\documentclass[aps,11pt, a4paper,notitlepage,groupedaddress, nofootinbib]{revtex4-1}
\pdfoutput=1

\usepackage[top=2.9cm, bottom=2.1cm, left=2cm, right=2cm]{geometry}
\usepackage{amsfonts,amssymb,amsmath}
\usepackage{graphicx}
\usepackage{booktabs}
\usepackage{theorem}
\usepackage{array}
\usepackage{colordvi}
\usepackage{xcolor}
\usepackage{subfigure}
\usepackage{fancyhdr}
\usepackage{epstopdf}
\usepackage{slashed}
\usepackage{hyperref}
\hypersetup{citecolor=green}
\usepackage{youngtab}

\newcolumntype{C}{>{$}c<{$}}
\AtBeginDocument{
\heavyrulewidth=.08em
\lightrulewidth=.05em
\cmidrulewidth=.03em
\belowrulesep=.65ex
\belowbottomsep=0pt
\aboverulesep=.4ex
\abovetopsep=0pt
\cmidrulesep=\doublerulesep
\cmidrulekern=.5em
\defaultaddspace=.5em
}

    \newcommand{\be}{\begin{equation}}
    \newcommand{\ee}{\end{equation}}
    \newcommand{\ba}{\begin{eqnarray}}
    \newcommand{\ea}{\end{eqnarray}}
    \newcommand{\eite}{\end{itemize}}
    \newcommand{\bite}{\begin{itemize}}
    \newcommand{\betabeta}{\mbox{$(\beta \beta)_{0 \nu}  $}}
    \def\ltap{\ \raisebox{-.4ex}{\rlap{$\sim$}} \raisebox{.4ex}{$<$}\ }
    \def\gtap{\ \raisebox{-.4ex}{\rlap{$\sim$}} \raisebox{.4ex}{$>$}\ }
    \newcommand{\meff}{\mbox{$\left|\langle\,\!  m \! \,\rangle\right| \ $}}
    \newcommand{\mefff}{\mbox{$ \langle\, \! m \! \,\rangle $}}

    \newcommand{\g}[1]{\ensuremath{\gamma#1}} 

    \newcommand{\al}{\alpha}
    
\newcommand{\La}[1]{\mathcal#1}

\begin{document}

\title{The nature of massive    neutrinos and Multiple  Mechanisms\\ in Neutrinoless Double Beta Decay
\footnote{Invited contribution written for the INFN Fubini Prize, year 2013/2014}}
\author{Aurora Meroni}                     
%
%
\email{meroni@cp3.dias.sdu.dk}
\affiliation{{CP}$^{ \bf 3}${-Origins} \& the Danish Institute for Advanced Study {\rm{Danish IAS}},  University of Southern Denmark, Campusvej 55, DK-5230 Odense M, Denmark.}
%
\begin{abstract}
Determining the nature ---Dirac or Majorana--- 
of massive neutrinos is one of the most pressing and challenging problems in the field of neutrino physics.
We discuss how one can possibly extract information on the couplings, if any, which might be involved 
in $\betabeta$-decay using a multi-isotope approach.  
We investigate as well the potential of combining data on the half-lives of nuclides which largely different Nuclear Matrix Elements such as $^{136}Xe$ and of one or more of the four nuclei $^{76}$Ge,  $^{82}$Se,  $^{100}$Mo and  $^{130}$Te, for discriminating between different pairs of non-interfering or interfering mechanisms of \betabeta-decay.
The case studies do not extend to the evaluation of the theoretical uncertainties of the results, due to the nuclear matrix elements calculations and other causes.
\end{abstract} 
%

%
\maketitle

\section{Are neutrinos Majorana particles?}

  All  neutrino oscillation data can be described
within the reference 3-flavour neutrino mixing scheme,
with 3 light neutrinos $\nu_j$, having masses $m_j \ltap 1$ eV, $j=1,2,3$
(see, e.g., \cite{pdg}). These data allowed to determine, with a relatively
high precision (see  \cite{Capozzi:2013csa, Gonzalez-Garcia:2014bfa} for  recent global fit analysis),
the parameters which drive the observed solar,
atmospheric, reactor and accelerator flavour
neutrino oscillations  ---the three neutrino mixing angles $\theta_{12}$, $\theta_{23}$ and $\theta_{13}$
of the standard parametrisation of the Pontecorvo, Maki,
Nakagawa and Sakata (PMNS) neutrino mixing matrix $U$ and the two
neutrino mass squared differences $\Delta m^2_{21}$ and
$\Delta m^2_{31}$ (or $\Delta m^2_{32}$).
\\
Despite the great successes of the last decades, we still have no clue about the nature of massive neutrinos, which could be Dirac or Majorana.
\\
The Majorana nature of  neutrinos
manifests itself in the existence of processes in which the total
lepton charge $L$ changes by two units: $K^+\rightarrow \pi^- +
\mu^+ + \mu^+$, $\mu^- + (A,Z)\rightarrow \mu^+ + (A,Z-2)$, etc.
Extensive studies have shown that the only feasible experiments
having the potential of establishing that the neutrinos are
Majorana particles are at present the experiments searching for
neutrinoless double beta \betabeta-decay (see e.g. for a review \cite{Vergados:2012xy}): 

\be (A,Z) \rightarrow (A,Z+2) + e^- + e^-.\label{eq:2bonu}\ee
Under the assumptions of 3-$\nu$ mixing, with neutrinos $\nu_j$
being Majorana particles, and of \betabeta-decay generated only by
the ($V-A$) charged current weak interaction via the exchange of the
three Majorana neutrinos $\nu_j$, the \betabeta-decay amplitude of interest has the form (see,
e.g.\cite{Rodejohann:2011mu,Bilenky2001bpp}):
\begin{equation}
\mathcal{A}(\beta\beta)_{0\nu} = \mefff \, M(A,Z)\,,
\end{equation}
%
where $M(A,Z)$ is the nuclear matrix element (NME) of the decay in eq. \eqref{eq:2bonu} which does not depend on the
neutrino mass and mixing parameters, and
\begin{equation}
\meff = \left| m_1\, |U_{\mathrm{e} 1}|^2
+ m_2\, |U_{\mathrm{e} 2}|^2\, e^{i\alpha_{21}}   
 + m_3 |U_{\mathrm{e} 3}|^2e^{i\alpha_{31}} \right|\,,
\label{effmass2}
\end{equation}
%
\noindent is the \betabeta-decay effective Majorana mass.
In eq. (\ref{effmass2}), $U_{ej}$, $j=1,2,3$, are the elements of the first
row of the PMNS matrix $U$, and
$\alpha_{21}$ and $\alpha_{31}$ are
the two Majorana CP violation (CPV) phases contained in $U$. In
the standard parametrization of the PMNS matrix $U$ (see, e.g.,
\cite{pdg}), the phase $\alpha_{31}$ in eq. (\ref{effmass2})
must be replaced by $(\alpha_{31} - 2\delta)$, $\delta$ being the
Dirac CPV phase present in $U$, and $|U_{\mathrm{e}1}| =
\cos\theta_{12}\cos\theta_{13}$, $|U_{\mathrm{e}2}| =
\sin\theta_{12}\cos\theta_{13}$, $|U_{\mathrm{e}3}| =
\sin\theta_{13}$.
\\
We recall that the predictions for  $\meff$
depend on the type of the neutrino mass spectrum
\cite{Bilenky2001bpp,Pascoli:2002xq, Pascoli:2003ke}. As is well known,
depending on the sign of $\Delta m^2_{31(32)}$,
which cannot be determined from
the presently available
neutrino oscillation data,
two types of neutrino mass spectrum are possible:\\
{\it i)  normal ordering (NO)}:
$m_1 < m_2 < m_3$,
$\Delta m^2_{31} >0$,
$\Delta m^2_{21} > 0$,
$m_{2(3)} = (m_1^2 + \Delta m^2_{21(31)})^{1/2}$; \\~~
{\it ii) inverted ordering (IO)}:
$m_3 < m_1 < m_2$,
$\Delta m^2_{32}< 0$,
$\Delta m^2_{21} > 0$,
$m_{2} = (m_3^2 + \Delta m^2_{32})^{1/2}$ and
$m_{1} = (m_3^2 + \Delta m^2_{32} - \Delta m^2_{21})^{1/2}$.\\
Depending on the value of the lightest neutrino mass,
$m_{min}$, the neutrino mass spectrum can be ($j=1,2,3$):\\
%
{\it a) Normal Hierarchical (NH)}:
$m_1 \ll m_2 < m_3$, $m_2 \cong (\Delta m^2_{21})^{1/2}
\cong 8.68 \times 10^{-3}$ eV,
$m_3 \cong (\Delta m^2_{31})^{1/2} 
\cong 4.97\times 10^{-2}~{\rm eV}$ or  \\
%
{\it b) Inverted Hierarchical (IH)}: $m_3 \ll m_1 < m_2$,
with $m_{1,2} \cong |\Delta m^2_{32}|^{1/2} \cong 4.97\times 10^{-2}$ 
eV
or  \\
%
{\it c) Quasi-Degenerate (QD)}: $m_1 \cong m_2 \cong m_3 \cong m_0$,
$m_j^2 \gg |\Delta m^2_{31(32)}|$ and $m_0 \gtap 0.10$ eV.
%
%
%
%

\begin{figure}[h!]
  \begin{center}
 \subfigure
 {\includegraphics[width=6cm]{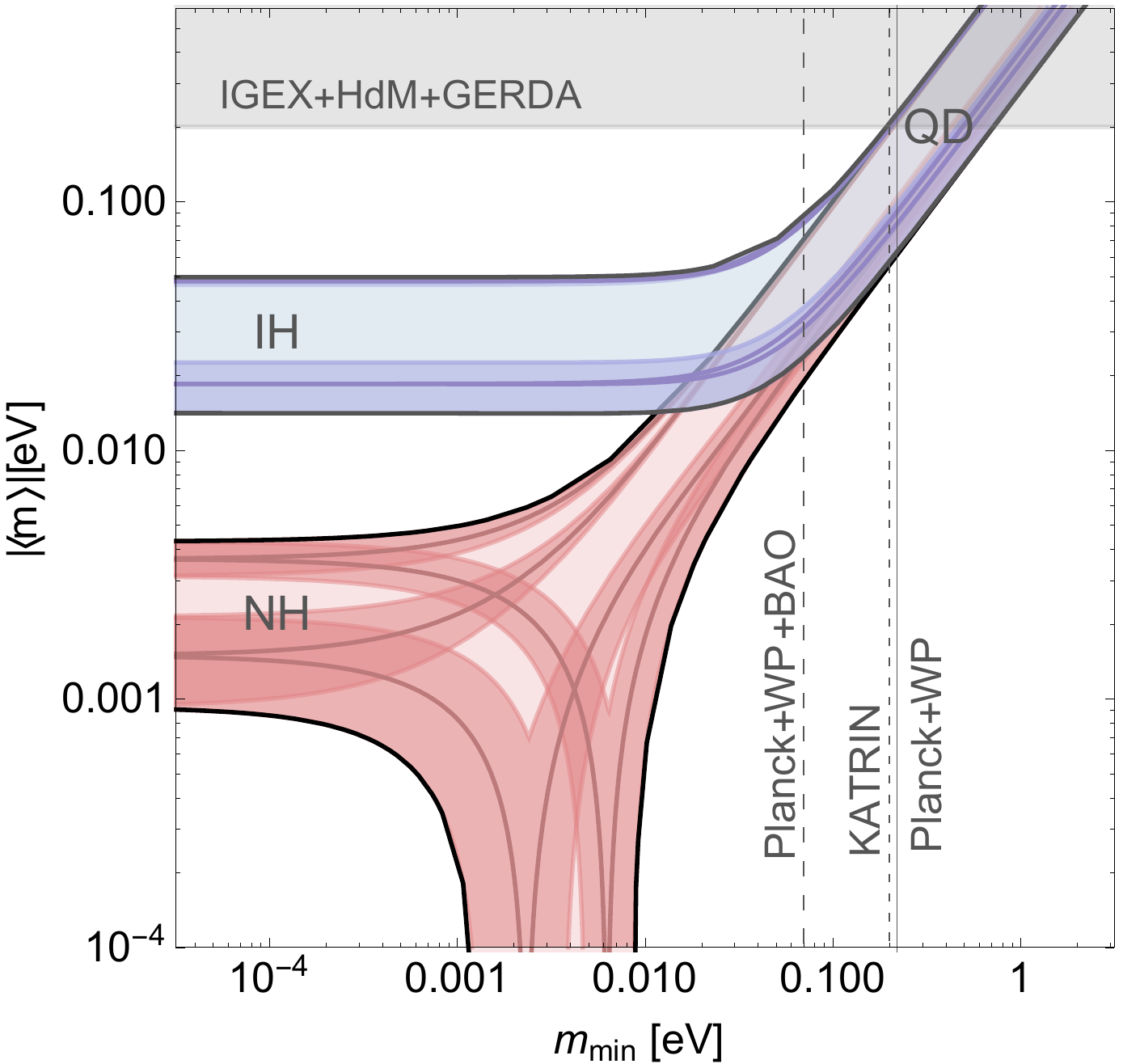}}
 \vspace{5mm}
 \subfigure
   {\includegraphics[width=6cm]{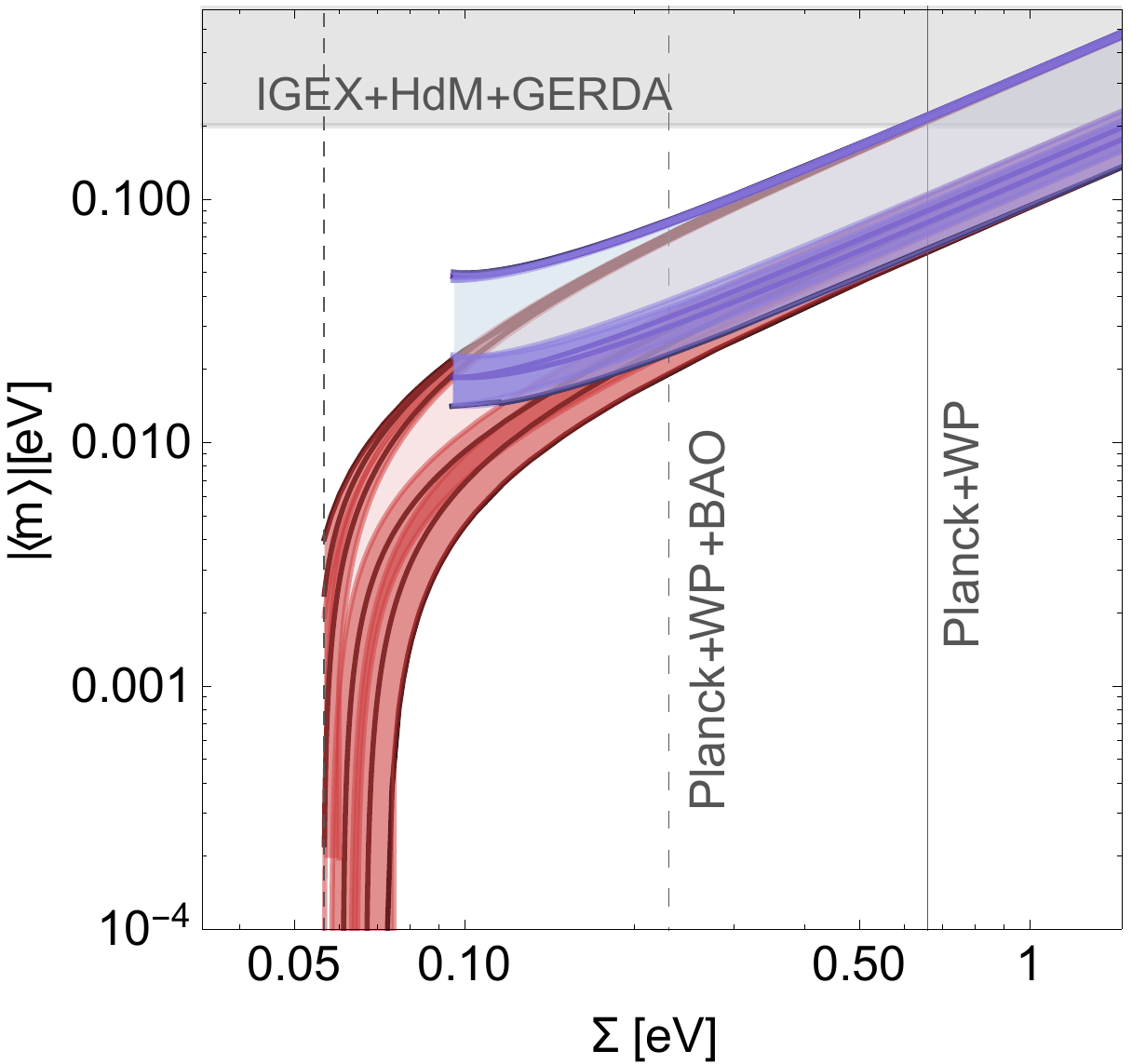}}
     \end{center}
\vspace{-1.0cm}
    \caption{\label{fig:meff}  Left Panel: The value of \meff as function
    of $m_{min}$. The dotted line corresponds to the 
    expected sensitivity of the future KATRIN $\beta$-decay experiment \cite{Eitel:2005hg}.
       Right Panel: Values of \meff as function of the sum of the neutrino masses $\Sigma$.  In 
       both panels  the light (strong) shaded regions indicate respectively  the
     $3\sigma$ allowed  CP-(non)conserving values of \meff. 
 In both plots  the vertical solid and dashed  lines are
    the constraints on $\Sigma$ obtained by the Planck Collaboration \cite{Ade:2013zuv}.  The grey exclusion band 
   is delimited by a value of $\meff=0.2$ eV,  obtained using the 90 \% C.L. limit on the half-life of $^{76}$Ge reported in 
    \cite{Agostini:2013mzu}.
}
\end{figure}

If the mass of the lightest neutrino would turn out to be extremely small, say $m_{min}\lesssim 10^{-3}$,
using the data on the neutrino oscillation
parameters one finds  that
\cite{Pascoli:2002xq, Pascoli:2003ke}
in the case of NH one has $\meff \lesssim
0.005$ eV, while if the spectrum is IH,
$0.01~{\rm eV} \lesssim \meff \lesssim 0.05$ eV 
(see left panel in Fig. \ref{fig:meff}). A larger value of
$\meff$ up to approximately 0.5 eV is possible if the light neutrino
mass spectrum is with partial hierarchy or is of quasi-degenerate
type. In the latter case $\meff$ can be close to the existing upper
limits. 

%

\begin{table}[h]\centering
\renewcommand{\tabcolsep}{1.5mm}
 \renewcommand{\arraystretch}{.7}
{
\begin{tabular}{cllcc}
\toprule
Isotope & $T_{1/2}^{0\nu}$ [10$^{25}$ yr]  & Experiment  & \meff [eV] \\[1mm]
\midrule
$^{136}$Xe &  $>1.6  $& \small{EXO-200}     & $0.16-0.30$   \\[2mm]
$^{136}$Xe &  $>1.9  $& \small{KamLAND-ZEN} & $0.14-0.28$   \\[2mm]
$^{76}$Ge &  $>2.1 $& \small{GERDA} & $0.23-0.54$  \\[2mm]
$^{76}$Ge &  $>3.0 $& \small{GERDA+IGEX+HdM} & $0.19-0.45$\\[2mm]
\bottomrule
\end{tabular}}\caption{\label{tab:limits}The experimental lower limits at 90\% C.L. on the \betabeta-decay
half-lives of different isotopes  and the corresponding lower and upper limits on $\meff$ computed with NMEs calculated in the framework of different 
approaches (a summary table is given  in \cite{Vergados:2012xy}).
}
\end{table}

 The most stringent upper limits on $\meff$ were set by the
IGEX ($^{76}$Ge), 
CUORICINO ($^{130}$Te),  
NEMO3 ($^{100}$Mo) 
and more recently by
 EXO-200 , KamLAND-ZEN ($^{136}$Xe) and GERDA ($^{76}$Ge) 
experiments
(see e.g. \cite{Piquemal:2013uaa} for a summary). 
A lower limit on the half-life of $^{76}$Ge,
T$_{1/2}^{0\nu} > 1.9\times 10^{25}$~yr (90\%~C.L.),
was found in the Heidelberg-Moscow $^{76}$Ge
experiment (HdM) ~\cite{KlapdorKleingrothaus:2000sn}.
Further a positive \betabeta-decay signal at $> 3\sigma$,
corresponding to T$_{1/2}^{0\nu} = (0.69 - 4.18)\times
10^{25}$~yr (99.73\%~C.L.) and implying $\meff = (0.1 -
0.9)~{\rm eV}$, was claimed to have been observed in
\cite{KlapdorKleingrothaus:2001ke}, and a later analysis reported
evidence for \betabeta-decay at 6$\sigma$
corresponding to $\meff = 0.32 \pm 0.03$~eV~\cite{01-Klap04}.
More recently, a large number of projects, or already running
experiments, have aimed at a sensitivity of $\meff \sim (0.01 - 0.05)$ eV,
i.e., to probe the range of  $\meff$
corresponding to IH  mass spectrum \cite{Piquemal:2013uaa}:
CUORE ($^{130}$Te), GERDA ($^{76}$Ge), SuperNEMO, EXO ($^{136}$Xe), MAJORANA
($^{76}$Ge), MOON ($^{100}$Mo), COBRA ($^{116}$Cd),
XMASS ($^{136}$Xe), CANDLES ($^{48}$Ca), KamLAND-Zen ($^{136}$Xe),
SNO+ ($^{150}$Nd), etc.
%
%
Specifically among these, GERDA (GERmanium Detector Array) at the Gran Sasso
Laboratory (Italy)($^{76}$Ge),  EXO-200 (Enriched Xenon Observatory) in New Mexico ($^{136}$Xe)
and KamLAND-Zen in Japan ($^{136}$Xe) are operational and  they
have been able to  test the claim in \cite{01-Klap04}. In 2012 EXO-200 has obtained a lower limit on the half-life
of $^{136}$Xe \cite{Auger:2012ar},
\be 
\mbox{$T_{1/2}^{0\nu} (^{136}$Xe)}>1.6\times 10^{25}\rm{yr \,\,at \,90\%\,\, C.L.},
\ee
 while later the experiment KamLAND-Zen reported the lower bound
 \cite{Gando:2012zm};
\be 
\mbox{$T_{1/2}^{0\nu}(^{136}$Xe)}>1.9\times 10^{25}\rm{yr \,\,at \,90\%\,\, C.L.}. 
\ee
%
%
In July 2013 also the  GERDA  collaboration reported the
results from Phase I   for \betabeta decay of the isotope $^{76}$Ge.
No signal was observed and a lower limit has been
set for $T_{1/2}^{0\nu}(^{76}$Ge) \cite{Agostini:2013mzu}:
\be 
\mbox{$T_{1/2}^{0\nu}(^{76}$Ge)}>2.1\times 10^{25}\rm{yr \,\,at \,90\%\,\, C.L.}, 
\ee
%
 which disprove, together with the other experimental results mentioned above,  the claim in \cite{01-Klap04}.
The GERDA collaboration reported also a combined limit using 
the bounds obtained by the
HdM and IGEX experiments, that reads:
\be 
\mbox{$T_{1/2}^{0\nu}(^{76}$Ge)}>3.0\times 10^{25}\rm{yr \,\,at \,90\%\,\, C.L.}.
\ee
%
\\
In Table \ref{tab:limits} we present the constraints on 
\meff obtained by the  \betabeta-decay 
experimental bounds mentioned above.
In order to extract these constraints,  we use as illustrative example, 
the sets of NMEs presented  in Table 3 in \cite{Vergados:2012xy} 
(see references therein for additional references on NMEs models) 
for the decays of the
nuclides  of interest, $^{76}$Ge and $^{136}$Xe.
%
%
%
%
%
%

In Fig. \ref{fig:meff} we show the predictions on \meff as function of $m_{min}$ (left panel) and as function of the sum of the neutrino masses $\Sigma$ (right panel). 
We combined all the
available data i.e we used  the best fit values and the 3$\sigma$ uncertainty of one of the  most
recent global fit analysis on the neutrino oscillation parameters
\cite{Gonzalez-Garcia:2014bfa}, the results coming from \betabeta-decays searches and the cosmological constraints from the Planck Collaboration \cite{Ade:2013zuv}. The latter
comes from  the CMB data   which  yields  an upper limit on the sum of the neutrino mass 
$\Sigma < 0.66 eV$ at 95\% C.L. (referred in Fig.  \ref{fig:meff}  as  ``Planck + WP"), and the CMB data combined with the Barion Acoustic Oscillation data (BAO), 
for which $\Sigma < 0.23 eV$ at 95\% C.L. (referred as ``Planck + WP + BAO").
In Fig. \ref{fig:meff}  we report also the expected sensitivity of 
the KArlsruhe TRItium Neutrino experiment (KATRIN) on
the absolute scale of neutrino masses
i.e.   $m_{\bar\nu_e} \sim 0.2$ eV 
\cite{pdg, Eitel:2005hg}, which is
expected to start the data taking in late 2015.
\\
%
From the left panel of Fig. \ref{fig:meff}, one can realize that,
 for $m_{min} \lesssim 10^{-3}$ eV, in the case of NH spectrum, $max\meff$  is considerably
smaller than $min\meff$ for the IH spectrum. This opens the
possibility to obtain information about the neutrino mass pattern
from a measurement of \meff$\neq0$ if $m_{min} \lesssim 10^{-3}$ eV. More specifically, in the case $m_{min}\lesssim 10^{-3}$ eV, a positive result in the future generation of \betabeta-decay experiments with
$\meff \gtrsim 0.01$ eV, will imply that the NH spectrum is excluded (barring possible destructive interfering effects generated by new physics). 
If instead  future searches will show that $\meff \lesssim 0.01$ eV, both the IH and QD spectrum
will be ruled out for massive Majorana neutrinos. If in addition, it
is established from oscillation experiments that $\Delta m^2_{31(32)}<0$,
then one would deduce that either  the  neutrinos
are Dirac fermions, or they are Majorana particles and
there are additional contributions to \betabeta-decay amplitude
which  interfere destructively \cite{Pascoli:2007qh}.
\\
Summarizing,  the studies on \betabeta-decay and a measurement of a
nonzero value of $\meff \geq$ (of a few $10^{-2}$ eV) could:
\begin{itemize}
  \item  establish the Majorana nature of massive neutrinos;
    \item give information on the type of neutrino mass spectrum.
  More specifically, a measured value of $\meff \sim$ few $10^{-2}$
  eV can provide unique constraints on, or even can
  allow one to determine, the type of neutrino mass spectrum if
  neutrinos $\nu_i$ are Majorana particles \cite{Pascoli:2002xq,Pascoli:2003ke, Bilenky2001bpp,Bilenky:2001xq};
%
  \item provide also unique information on the absolute scale of
  neutrino masses or on the lightest neutrino mass (see e.g. \cite{PPW02ppw});
  \item with additional information from other sources ($^{3}H$ $\beta$-decay experiments or cosmological and astrophysical data
  considerations) on the absolute neutrino mass scale, the
  \betabeta-decay experiments can provide unique information on the
  Majorana CP-violation phases $\al_{31}$ and/or $\al_{21}$ \cite{Bilenky:1996cb,Pascoli:2005zb,Bilenky2001bpp}. 
\end{itemize}
%
%
%
%
If the $\betabeta$-decay will be observed, 
the corresponding data will be used 
to constrain or even determining the possible
mechanism(s) generating the decay. Indeed,  the observation
of the $\betabeta$-decay would not  guarantee
that the dominant mechanism inducing the decay is the
light Majorana neutrino exchange \cite{Duerr:2011zd} (see also \cite{Barry:2010en, Meroni:2014tba,Girardi:2013zra,Giunti:2015kza} for an example of how additional sterile neutrinos states   affect the decay).
The results of the \betabeta-decay searches will play a 
very important role in testing and constraining 
i) theories of 
neutrino mass generation
predicting massive Majorana neutrinos, and 
ii) the existence of new
$|\Delta L| = 2$ couplings in the effective weak interaction
Lagrangian, which could induce the decay. 
The existence of such couplings would have tremendous impact
from the model-building point of view.

\section{Multiple mechanisms in \betabeta-decay }\label{intro}

An observation of $\betabeta$-decay would
imply that the total lepton charge $L$ is not conserved. This of
course implies  that the massive neutrinos get a Majorana mass
\cite{SchVal82, 1984Takasugi} and therefore are Majorana particles
(see, e.g. \cite{Bilenky1987bp}). However, the latter does not
guarantee that the dominant mechanism inducing the $\betabeta$-decay
is the light Majorana neutrino exchange (that we will call the
``standard'' mechanism of the $\betabeta$-decay) since the Majorana
mass thus generated is exceedingly small. The $\betabeta$-decay can
well be due to the existence of interactions which do not conserve
the total lepton charge $L$, specifically $\Delta L = \pm 2$. A number of such
interactions have been proposed in the literature: heavy Majorana
neutrinos coupled to the electron in the $V-A$ charged current weak
interaction Lagrangian, supersymmetric (SUSY) theories with
$R$-parity breaking terms which do not conserve the total lepton
charge $L$, $L$-nonconserving couplings in the Left-Right symmetric
theories, etc. At present we do not have evidence for the existence
of $\Delta L \neq 0$ terms in the Lagrangian describing the particle
interactions. Nevertheless, such terms can exist and they can be
operative in the $\betabeta$-decay. Moreover, it is impossible to
exclude the hypothesis that, if observed, the $\betabeta$-decay is
triggered by more than one competing mechanisms.

The possibility of several different mechanisms contributing to the
$\betabeta$-decay amplitude was considered  in \cite{FSV10MM}
assuming that the corresponding $\Delta L = \pm 2$ couplings are CP
conserving. 

The analysis presented here  is a natural
continuation of the study performed in \cite{FSV10MM}. We consider
the possibility of several different mechanisms contributing to the
$\betabeta$-decay amplitude in the general case of CP
nonconservation: light Majorana neutrino exchange, heavy left-handed
(LH) and heavy right-handed (RH) Majorana neutrino exchanges, lepton
charge non-conserving couplings in SUSY theories with $R$-parity
breaking. These
different mechanisms can interfere only if the electron current
structure coincides and hence it can be factorized. If, on the
contrary, these are not-interfering mechanisms, i.e., the electron
currents have different chiralities, then the interference term is
suppressed by a factor which depends on the considered nucleus.
\cite{Halprin:1983ez}. If the $\betabeta$-decay is induced by, e.g., two
``non-interfering'' mechanisms (e.g. light Majorana neutrino and heavy RH
Majorana neutrino exchanges), one can determine the absolute values
of the two fundamental parameters, characterizing these mechanisms,
from data on the half-lives of two nuclear isotopes. In the case
when two ``interfering'' mechanisms are responsible for the
$\betabeta$-decay, the absolute values of the two relevant
parameters and the interference term can be uniquely determined from
data on the half-lives of three nuclei. We present in section \ref{multiple} illustrative examples of determination
of the relevant fundamental parameters and of possible tests of the
hypothesis that more than one mechanism is responsible for the
$\betabeta$-decay, using as input hypothetical half-lives of
$^{76}$Ge, $^{130}$Te and $^{100}$Mo and considering two ``noninterfering'' and two ``interfering''
mechanisms,
namely, the light Majorana neutrino and the heavy RH Majorana
neutrino exchanges, and the light Majorana neutrino and the dominant
gluino exchanges. The effects of the
uncertainties in the values of the nuclear matrix elements (NMEs) on
the results of the indicated analyzes are also discussed and
illustrated.\\
The method considered here can be generalized to the case of more
than two $\betabeta$-decay mechanisms. It has also the advantage
that it allows to treat the cases of CP conserving and CP
nonconserving couplings generating the $\betabeta$-decay in a unique
way.
In section \ref{multiple2} we will investigate also the potential of
combining data on one or more of
the five nuclei $^{76}$Ge, $^{82}$Se, $^{100}$Mo,  $^{130}$Te and  $^{136}$Xe, for
discriminating between different pairs of non-interfering or
interfering mechanisms of $\betabeta$-decay.

\section{Possible $\Delta L=2$ coupligs in $(\beta \beta)_{0 \nu}$-Decay}

The \betabeta-decay is allowed by a number of models,
from the standard mechanism of light Majorana neutrino exchange
 to those such as Left-Right Symmetry
\cite{Mohapatra:1980yp,Tello:2010am} or R-parity violating
Supersymmetry (SUSY) \cite{Hirsch:1995ek}. These mechanisms might
trigger $\betabeta$-decay individually or together. 
The $\betabeta$-decay half-life for a certain
nucleus can therefore be written as function of some lepton number
violating (LNV) parameters, each of them connected with a different
mechanism $i$:
\be [ T^{0\nu}_{1/2}]^{-1} = G^{0\nu}(E, Z) |\sum_i \eta_i^{LNV}
    {\cal M}^{0\nu}_{i}    |^2 \ee
where $G^{0\nu}(E,Z)$ and $ {\cal M}^{0\nu}_{i} $
are, respectively, the  phase-space factor ($E_0$ is the energy
release) and the NMEs of the decay.  The values of the phase space factor will be taken from  \cite{Meroni:2012qf}. Further, The NMEs
depends on the mechanism generating the decay.
Here, following the analysis in \cite{Faessler:2011qw, Meroni:2012qf},
we will adopt 
the Self-consistent
Renormalized Quasiparticle Random Phase Approximation (SRQRPA)
\cite{Delion:1997vr,Simkovic:2008cu}. In this approach, the
particle-particle strength parameter $g_{pp}$ of the SRQRPA
\cite{Rodin1, Rodin:2007fz, Simkovic:2007vu} is fixed by 
the recent data on the two-neutrino
double beta decays of EXO-200. In the calculation of the
\betabeta-decay NMEs were considered the two-nucleon short-range
correlations derived from same potential as residual interactions,
namely from the Argonne or CD-Bonn potentials \cite{Simkovic:2009} and two
values of the axial-vector constant are used, $g_A=1.0, 1.25$.
The  numerical values  used in the following analysis are taken from \cite{Meroni:2012qf}.
Let us make now some concrete examples on possible \betabeta decay parameters.
In the case of the light Majorana neutrino exchange mechanism of \betabeta-decay, we can define a LNV parameter which is given by:
\be \eta_\nu = \frac{\langle m \rangle}{m_e}\ee
where $m_e$ is the electron mass. 
Next, we assume that the neutrino mass spectrum includes,
in addition to the three light Majorana neutrinos,
heavy Majorana states $N_k$ with masses
$M_k$ much larger than the typical energy scale of the
$\betabeta$-decay, $M_k \gg 100$ MeV;
we will consider the case of $M_k \gtap 10$ GeV.
Such a possibility arises
if the weak interaction Lagrangian includes
right-handed (RH) sterile neutrino fields which
couple to the LH flavour neutrino fields
via the neutrino Yukawa coupling and possess a
Majorana mass term.
The heavy Majorana neutrinos $N_k$
can mediate the $\betabeta$-decay
similar to the light Majorana neutrinos
via the $V-A$ charged current weak interaction.
The difference between the two mechanisms is that,
unlike in the light Majorana neutrino exchange
which leads to a long range inter-nucleon interactions,
in the case of $M_k \gtap 10$ GeV
of interest the momentum dependence
of the heavy Majorana neutrino propagators
can be neglected (i.e., the $N_k$ propagators
can be contracted to points)
and, as a consequence, the corresponding
effective nucleon transition operators are local.
The LNV parameter in the case when the
$\betabeta$-decay
is generated by the $(V-A)$ CC
weak interaction due to the
exchange of $N_k$ can be written as:
\begin{eqnarray}
\eta_{_L}
~&=& ~ \sum^{heavy}_k~ U_{ek}^2
\frac{m_p}{M_k}\,,
\label{etaL}
\end{eqnarray}
%
where $m_p$ is the proton mass and  $U_{ek}$
is the element of the neutrino mixing matrix
through which $N_k$ couples to the electron in the
weak charged lepton current.
 If the weak interaction Lagrangian contains also
$(V+A)$ (i.e., right-handed (RH)) charged currents
coupled to a RH charged weak boson $W_R$,
as,
\be \La{L}_{L+R} = \frac{g}{2\sqrt{2}}[(\bar e
\g{_\alpha}(1-\g{_5})\nu_{e})W_{\mu L}^{-}+(\bar e
\g{_\alpha}(1+\g{_5})\nu_{e})W_{\mu R}^{-}]\ee where
$\nu_{eR}=\sum_k V_{ek}N_{kR}$, $C\bar N^T_k= \xi N_k$. Here $V_{e
k}$ are the elements of a mixing matrix by which $N_k$ couple to the
electron in the $(V+A)$ charged lepton current,
$M_{L}^W$ is the mass of the Standard Model charged weak boson,  $M_{L}^W
\cong 82$ GeV, and $M_{R}^W$ is the mass of $W_R$. It follows from
the existing data that \cite{Mohapatra:1980yp, Tello:2010am} $W_R \gtap 2.5$ TeV.
For instance, in the $SU(2)_L\times SU(2)_R\times U(1)$ theories we
can have also a contribution to the
$\betabeta$-decay amplitude generated
by the exchange of virtual $N_k$
coupled to the electron in
the hypothetical $(V+A)$ CC
part of the weak interaction Lagrangian.
In this case the corresponding LNV parameter
can be written as:
\begin{eqnarray}
\eta^{}_{_R}
~&=& ~ \left (\dfrac{M^W_L}{M^W_{R}}\right )^{4}\,\sum^{heavy}_k~ V_{e k}^2
\frac{m_p}{M_k}.
\label{etaR}
\end{eqnarray}
If CP invariance does not hold, which
we will assume to be the case in what
follows, $U_{ek}$ and $V_{ek}$
will contain physical CP violating
phases at least for some $k$ and thus
the parameters $\eta_{_L}$
and $\eta_{_R}$ will not be real.
 \begin{figure}
  \begin{center}
   \includegraphics[width=6.5cm]{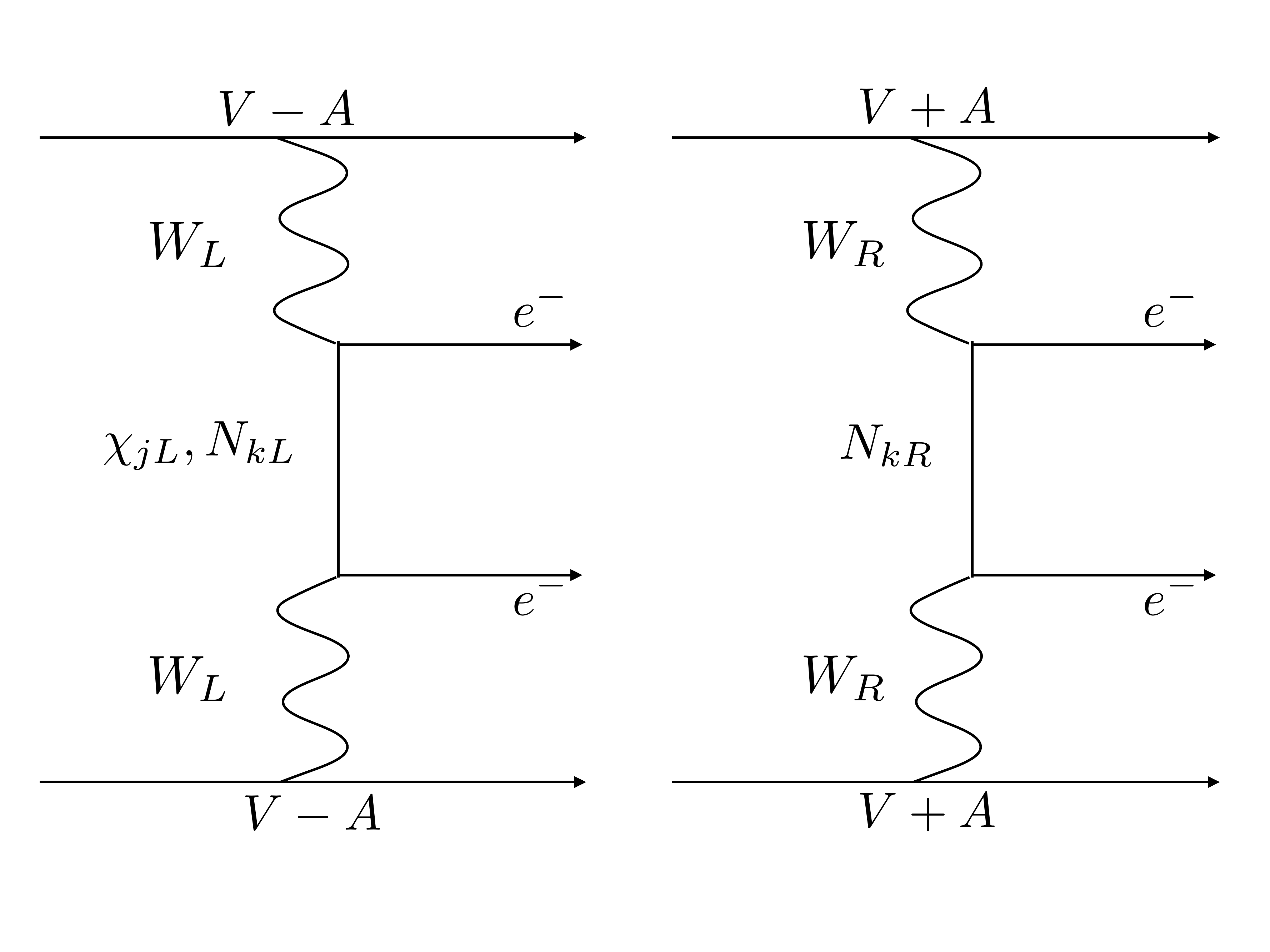}
   \end{center}
\vspace{-20pt}\caption{\label{Feyn1} Left Panel: Feynman diagrams for the $\betabeta$-decay,
generated by the light or heavy (LH) Majorana neutrino exchange (respectively $\chi_{jL}$, and $N_{kL}$).
Right Panel:  the heavy (RH) Majorana neutrino exchange.}
\end{figure}
 As can be shown,
the NMEs
corresponding to the two mechanisms of
$\betabeta$-decay with exchange of heavy
Majorana neutrinos $N_k$, described in the
present section, are the same
and are  given in \cite{Simkovic:1999re}.
We will denote them by ${\cal M}^{0\nu}_{_N}$.
Finally, it is important to note that
the current factor in the
$\betabeta$-decay amplitude describing
the two final state electrons,
has different forms in the cases of
$\betabeta$-decay mediated by  $(V-A)$ and
$(V+A)$ CC weak interactions \footnote{The procedure is the same defined in the
light neutrino exchange section. One has in this case
$\bar e \g{_\al} P_R \g{_\beta}  C \bar e^T A_{\al\beta}=
 \bar e P_L e^c A_{\al\beta}$  }, namely,
$\bar{e}(1 + \gamma_5)e^c \equiv
2\bar{e_L}\, (e^c)_R$
and  $\bar{e}(1 - \gamma_5)e^c
\equiv 2\bar{e_R}\,(e^c)_L $,
respectively, where $e^c = C(\bar{e})^{T}$,
$C$ being the charge conjugation  matrix.
The difference in the chiral structure
of the two currents leads to a relatively
strong suppression of the interference
between the terms in the
$\betabeta$-decay amplitude
involving the two different
electron current factors.

\section{Uncovering Multiple CP non-conserving Mechanisms}\label{multiple}

We are going to illustrate the possibility to get information
about the different LNV parameters when two or more mechanisms are
operative in  $\betabeta$-decay, analyzing the following two cases.
First we consider two competitive ``not-interfering'' mechanisms of
$\betabeta$-decay: light left-handed Majorana neutrino exchange and
heavy right-handed Majorana neutrino exchange. In this case the
interference term arising in the $\betabeta$-decay half-life from
the product of the contributions due to the two mechanisms in the
$\betabeta$-decay amplitude, is strongly suppressed \cite{Halprin:1983ez} as
a consequence of the different chiral structure of the final state
electron current in the two amplitudes. The latter leads to a
different phase-space factor for the interference term, which is
typically by a factor of 10 smaller than the standard one
(corresponding to the contribution to the $\betabeta$-decay
half-life of each of the two mechanisms). More specifically, the
suppression factors for $^{76}$Ge, $^{82}$Se,  $^{100}$Mo and
$^{130}$Te read, respectively \cite{Halprin:1983ez}: 0.13; 0.08; 0.075 and
0.10. It is particularly small for $^{48}$Ca: 0.04. In the analysis
which follows we will neglect the contribution of the interference
term in the  $\betabeta$-decay half-life. The effect of taking into
account the interference term on the results thus obtained, as our
numerical calculations have shown, does not exceed approximately
10\%.

  In the case of negligible
interference term, the inverse value of the $\betabeta$-decay
half-life for a given isotope (A,Z) is given by:
\be \frac{1}{T^{0\nu}_{1/2,i}G^{0\nu}_i(E, Z)} \cong |\eta_\nu|^2
|{\cal M}^{0\nu}_{i,\nu}|^2 + |\eta_{_R}|^2|{\cal M}^{0\nu}_{i,N }|^2\,,
\label{hl} \ee
%
where the index $i$ denotes the isotope. 
  In the second illustrative
case we consider $\betabeta$-decay triggered by two active and
``interfering'' mechanisms: the light Majorana neutrino exchange and
the LH Majorana heavy neutrino exchange. In this case, for a given nucleus, the inverse
of the $\betabeta$-decay half-life is given by:
\be \frac{1}{T^{0\nu}_{1/2,i}G^{0\nu}_i(E, Z) }=|\eta_\nu|^2
|{\cal M}^{0\nu}_{i,\nu}|^2 +
|\eta^{}_{_L}|^2|{\cal M}^{0\nu}_{i,N}|^2 + 2\cos\alpha
|{\cal M}^{0\nu}_{i,N}||{\cal M}^{0\nu}_{i,\nu}||\eta_\nu||\eta_{_L}|\,.
\label{hlint} \ee
%
Here $|\eta_{_L}|$ is the basic parameter of the LH heavy Neutrino
exchange mechanism defined in eq. (\ref{etaL}) and $\alpha$ is
the relative phase of $\eta_{_L}$ and $ \eta_\nu$. 

We will use in the following illustrative examples  how one
can extract information about $ |\eta^{LNV}|$, using
the lower limits on the \betabeta-decay half-lives of
$^{76}$Ge, $^{82}$Se and $^{100}$Mo, and of $^{130}$Te reported by
the Heidelberg-Moscow \cite{Baudis:1999xd},  NEMO3
\cite{Barabash:2010bd} and CUORICINO \cite{Arnaboldi08} experiments,
respectively, as well as as well the $^{76}$Ge half-life reported in \cite{KlapdorK:2006ff} (see also \cite{01-Klap04}):

\be
\begin{split}
T^{0\nu}_{1/2}(^{76}Ge)> 1.9\times 10^{25} y \mbox{ y \cite{Baudis:1999xd}} & \quad T^{0\nu}_{1/2}(^{82}Se) >  3.6\times
10^{23} \mbox{{ y}~
\cite{Barabash:2010bd}},
\\
T^{0\nu}_{1/2}(^{100}Mo) >  1.1\times 10^{24} \mbox{{ y}~
\cite{Barabash:2010bd}},&\quad T^{0\nu}_{1/2}(^{130}Te)
> 3.0\times 10^{24} \mbox{{ y}~\cite{Arnaboldi08}}\,.
\\
T^{0\nu}_{1/2}(^{76}Ge)= 2.23^{+0.44}_{-0.31}\times 10^{25}\mbox{{ y}~
\cite{KlapdorK:2006ff}}\,.
\end{split}
\label{limit}
\ee
%
 In the analysis which follows we will present numerical
results first for $g_A=1.25$ and using the NMEs calculated with the
large size single particle basis (``large basis'') and the Charge
Dependent Bonn (CD-Bonn) potential. Later results for $g_A=1.0$, as
well as for NMEs calculated with the Argonne potential, will also be
reported.

\subsection{\label{sec:anal} Two ``Non-Interfering'' Mechanisms}
%
In this case the solutions for the corresponding two LNV parameters
$|\eta_A|^2$ and $|\eta_B|^2$ obtained from data on the
$\betabeta$-decay half-lives of the two isotopes $(A_i,Z_i)$ and
$(A_j,Z_j)$, are given by:
\be
\begin{split}
|\eta_A|^2 =\frac{|{\cal M}^{0\nu}_{j,B }|^2/T_i G_i- |{\cal M}^{0\nu}_{i,B
}|^2/T_j G_j} {|{\cal M}^{0\nu}_{i,A}|^2|{\cal M}^{0\nu}_{j,B
}|^2-|{\cal M}^{0\nu}_{i,B}|^2 |{\cal M}^{0\nu}_{j,A }|^2},\quad
|\eta_B|^2=\frac{ |{\cal M}^{0\nu}_{i,A }|^2/T_j G_j - |{\cal M}^{0\nu}_{j,A
}|^2/T_i G_i} {|{\cal M}^{0\nu}_{i,A}|^2|{\cal M}^{0\nu}_{j,B }|^2-
|{\cal M}^{0\nu}_{i,B}|^2|{\cal M}^{0\nu}_{j,A }|^2}. \label{solnonint}
\end{split}
\ee
%
It follows from eq. (\ref{solnonint}) that   if one
of the two half-lives, say $T_i$, is fixed, the positivity
conditions $|\eta_A|^2 \geq 0$ and $|\eta_B|^2 \geq 0$ can be
satisfied only if $T_j$ lies in a specific ``positivity interval''.
Choosing for convenience always $A_j <A_i$ we get for the positivity
interval:
\be \frac{ G_i}{G_j} \frac{  |{\cal M}^{0\nu}_{i,B }|^2 } {
|{\cal M}^{0\nu}_{j,B }|^2} T^{exp}_{i~min}\leq    \frac{ G_i}{G_j}    \frac{  |{\cal M}^{0\nu}_{i,B }|^2 }{
|{\cal M}^{0\nu}_{j,B }|^2} T_i \leq T_j \leq \frac{ G_i}{G_j} \frac{
 |{\cal M}^{0\nu}_{i,A }|^2 }{ |{\cal M}^{0\nu}_{j,A }|^2} T_i\,,
\label{PosC} \ee
%
where we have used $|{\cal M}^{0\nu}_{i,A }|^2/|{\cal M}^{0\nu}_{j,A }|^2
> |{\cal M}^{0\nu}_{i,B }|^2 /|{\cal M}^{0\nu}_{j,B }|^2$ and the first
 inequality in eq. (\ref{PosC})  has been obtained considering 
 the lower bound of interest for the isotope $(A_i,Z_i)$, i.e., if
$T_i \geq T^{exp}_{i~min}$.
 If only one of the mechanism is present, e.g.  the light neutrino exchange mechanism,
and the NME are correctly calculated, the
$\betabeta$-decay effective Majorana mass (and $|\eta_\nu|^2$)
extracted from all three (or any number of) $\betabeta$-decay
isotopes must be the same (see, e.g.,  \cite{PPW02ppw,BiPet04}).
Similarly, if the heavy RH Majorana neutrino exchange gives the
dominant contribution, the extracted value $|\eta_{_R}|^2$ must be the
same for all three (or more) $\betabeta$-decay nuclei.
Assuming $\eta_A\equiv\eta_\nu$ and $\eta_B\equiv\eta_{_R}$ and using  the
values of the phase-space factors and the two relevant NME for ``CD-Bonn, large, $g_A=1.25(1.0) $'', we get  for example:
\be
0.15  \leq \frac{T(^{100}Mo)}{T(^{76}Ge)}
\leq 0.18 (0.17)\,,
0.17 \leq \frac{T(^{130}Te)}{T(^{76}Ge)}
\leq 0.22 (0.23)\,,
1.14 (1.16)\leq \frac{T(^{130}Te)}{T(^{100}Mo)}
\leq 1.24(1.30)\,. \label{CDBonn125} \ee
It is quite remarkable that the physical solutions are possible only
if the ratio of the half-lives of all the pairs of the three
isotopes considered take values in very narrow intervals. This
result is a consequence of the values of the phase space factors and
of the NME for the two mechanisms considered. In the case of the
Argonne potential, ``large basis'' and $g_A=1.25~(1.0)$ we get very similar results:
\be
0.15  \leq \frac{T(^{100}Mo)}{T(^{76}Ge)}
\leq 0.18\,, 0.18 \leq
\frac{T(^{130}Te)}{T(^{76}Ge)} \leq
0.24~(0.25)\,, 1.22 \leq
\frac{T(^{130}Te)}{T(^{100}Mo)} \leq
1.36~(1.42)\,. \label{Argonne12510} \ee
%
If it is experimentally established that any of the three ratios of
half-lives considered lies outside the interval of physical
solutions  of $|\eta_\nu|^2$ and  $|\eta_{_R}|^2$, obtained taking into
account all relevant uncertainties, one would be led to conclude
that the $\betabeta$-decay is not generated by the two mechanisms
under discussion. In order to show that the constraints given above
are indeed satisfied, the relevant ratios of $\betabeta$-decay
half-lives should be known with a remarkably small uncertainty (not
exceeding  approximately 5\% of the central values of the
intervals).

Obviously, given the half-life of one isotope, constraints similar
to those described above can be derived on the half-life of any
other isotope beyond those considered by us. Similar constraints can
be obtained in all cases of two ``non-interfering'' mechanisms
generating the $\betabeta$-decay. The predicted  intervals of
half-lives of the various  isotopes will differ, in general, for the
different pairs of  ``non-interfering''  mechanisms. However, as it is shown  \cite{Faessler:2011qw}, these
differences in the cases of the $\betabeta$-decay triggered
by the exchange of heavy Majorana neutrinos coupled to (V+A)
currents  and i) the gluino exchange mechanism, or ii)  the
squark-neutrino exchange mechanism,  are extremely  small. One of
the consequences of this feature of the different pairs of
``non-interfering'' mechanisms considered  by us is that if it will
be possible to rule out one of them as the cause of
$\betabeta$-decay, most likely one will  be able to rule out all
three of them. The set of constraints  under discussion will not be
valid, in general,    if the  $\betabeta$-decay is triggered by two
``interfering''  mechanisms with a non-negligible interference term,
or by more than two mechanisms with significant contributions to the
$\betabeta$-decay rates of the different nuclei.

\begin{table}
\centering \caption{\label{tab:table2} The predictions for the
half-life of a third nucleus $(A_3,Z_3)$, using as input in the
 equations for $|\eta_\nu|^2$ and $|\eta_{_R}|^2$, eq.
(\ref{solnonint}), the half-lives of two other nuclei $(A_1,Z_1)$ and
$(A_2,Z_2)$. The three nuclei used are $^{76}$Ge, $^{100}$Mo and
$^{130}$Te. The results shown are obtained for a fixed value of the
half-life of  $(A_1,Z_1)$ and assuming the half-life of  $(A_2,Z_2)$
to lie in a certain specific interval. The physical solutions for
$|\eta_\nu|^2$ and $|\eta_{_R}|^2$ and then used to derive predictions
for the half-life of the third nucleus  $(A_3,Z_3)$. The latter are
compared with the lower limits given in eq.  (\ref{limit}). The
results quoted are obtained for NMEs given in the columns
``CD-Bonn, large, $g_A=1.25$'' in  \cite{Meroni:2012qf}. One star
beside the isotope pair whose half-lives are used as input for the
system of equations (\ref{solnonint}), indicates predicted ranges of
half-lives of the nucleus $(A_3,Z_3)$ that are not compatible with
the lower bounds given in (\ref{limit}).}
\renewcommand{\arraystretch}{0.8}
{\footnotesize\tt
\begin{tabular}{|l|l|c|c|}
\toprule
Pair  &  T$^{0\nu}_{1/2}(A_1,Z_1)$[yr] & T$^{0\nu}_{1/2}[A_2,Z_2]$[yr]   &  Prediction on $[A_3,Z_3]$[yr] \\
\midrule
$^{76}Ge-^{100}Mo$  &  T$(Ge)= 2.23\cdot10^{25}$&  $3.23 \cdot10^{24}\leq T(Mo) \leq 3.97 \cdot10^{24}$   & $3.68 \cdot10^{24}\leq  T(Te) \leq 4.93 \cdot10^{24}$\\
$^{76}Ge-^{130}Te$  &   T$(Ge)=2.23\cdot10^{25}$&  $3.68 \cdot10^{24}\leq  T(Te) \leq 4.93 \cdot10^{24}$    & $3.23 \cdot10^{24}\leq T(Mo) \leq 3.97 \cdot10^{24}$\\
$^{76}Ge-^{100}Mo$  & T$(Ge)= 10^{26}$ & $1.45 \cdot10^{25}\leq
T(Mo)\leq 1.78 \cdot10^{25}$    & $1.65 \cdot10^{25}\leq  T(Te) \leq 2.21 \cdot10^{25}$  \\
$^{76}Ge-^{130}Te$  &  T$(Ge)= 10^{26}$ &  $1.65 \cdot10^{25}\leq
T(Te) \leq 2.21 \cdot10^{25}$   &$1.45 \cdot10^{25}\leq
T(Mo)\leq 1.78 \cdot10^{25}$ \\
$^{100}Mo-^{130}Te$ $\star$  &  T$(Mo)=5.8\cdot10^{23}$&  $6.61 \cdot10^{23}\leq T(Te)\leq 7.20 \cdot10^{23}$   &  $3.26 \cdot10^{24}\leq T(Ge) \leq4.00 \cdot10^{24}$\\
$^{100}Mo-^{130}Te$  &T$(Mo)=4\cdot10^{24}$& $ 4.56 \cdot10^{24}\leq T(Te) \leq 4.97 \cdot10^{24}$       & $ 2.25 \cdot10^{25}\leq T(Ge) \leq 2.76 \cdot10^{25}$\\
$^{100}Mo-^{130}Te$  & T$(Mo)=5.8\cdot10^{24}$&  $6.61\cdot10^{24}\leq T(Te) \leq 7.20 \cdot10^{24}$     &  $3.26\cdot10^{25}\leq T(Ge) \leq 4.00 \cdot10^{25}$\\
$^{100}Mo-^{130}Te$ $\star$ &  T$(Te)=3\cdot10^{24}$& $2.42
\cdot10^{24}\leq T(Mo) \leq 2.63 \cdot10^{24}$           & $1.36
\cdot10^{25}\leq T(Ge) \leq 1.82
\cdot10^{25}$\\
$^{100}Mo-^{130}Te$  &   T$(Te)=1.65\cdot10^{25}$&  $1.33 \cdot10^{25}\leq T(Mo) \leq 1.45 \cdot10^{25}$      &  $7.47 \cdot10^{25}\leq T(Ge) \leq 1.00 \cdot10^{26}$\\
$^{100}Mo-^{130}Te$  &   T$(Te)=3\cdot10^{25}$&  $2.42 \cdot10^{25}\leq T(Mo) \leq 2.63 \cdot10^{25}$         &  $1.36 \cdot10^{26}\leq T(Ge) \leq 1.82 \cdot10^{26}$\\
\bottomrule
\end{tabular}}
\end{table}
%
 We analyze next the possible solutions  for
different combinations of the half-lives of the following isotopes:
$^{76}$Ge, $^{100}$Mo and $^{130}$Te. Assuming the half-lives of two
isotopes to be known and using the physical solutions for
$|\eta_\nu|^2$ and $|\eta_{_R}|^2$ obtained using these half-lives, one
can obtain a prediction for the half-life of the third isotope. The
predicted half-life should satisfy the existing lower limits on it.
In the calculations the results of which are reported here, we fixed
the half-life of one of the two isotopes and assumed the second
half-life lies in an interval compatible with the existing
constraints. We used the value of T$^{0\nu}_{1/2}(^{76}$Ge$)$ and
values of  T$^{0\nu}_{1/2}(^{100}$Mo$)$ and T$^{0\nu}_{1/2}(^{130}$Te$)$
from the intervals given in (\ref{limit}). The system of two
equations is solved and the values of  $|\eta_\nu|^2> 0$ and
$|\eta_{_R}|^2> 0$ thus obtained were used to obtained predictions for
the half-life of the third isotope. The results for NMEs
corresponding to the case
 ``CD-Bonn, large, $g_A=1.25$''
 are given in Table \ref{tab:table2}. We
note that the experimental lower bounds quoted in eq. (\ref{limit})
have to be taken into account since they can further constrain the
range of allowed values of $|\eta_\nu|^2$ and $|\eta_{_R}|^2$. Indeed,
an inspection of the values in Table \ref{tab:table2} shows that not
all the ranges predicted for the third half-life using the solutions
obtained for $|\eta_{_R}|^2$ and $|\eta_{\nu}|^2$ are compatible with
the lower bounds on the half-live of the considered nuclear
isotopes, given in (\ref{limit}). In this case, some or all
``solution'' values of  $|\eta_{_R}|^2$ and/or $|\eta_{\nu}|^2$ are
ruled out. In Table \ref{tab:table2} these cases are marked by a
star.

 The results reported in Table \ref{tab:table2}
are stable with respect to variations of the NMEs. If we use the
NMEs corresponding to the case ``CD-Bonn, large, $g_A=1.0$'', the limits of the
intervals quoted in  Table \ref{tab:table2} change by $\pm
5\%$. If instead we use the NMEs corresponding to the Argonne
potential, ``large basis'' and $g_A=1.25$ ($g_A=1.0$), the indicated
limits change by $\pm 10\%$ ($\pm 14\%$).

\section{Largely Different NMEs and $\betabeta$-Decay}\label{multiple2}
  The observation of \betabeta-decay of several
different isotopes is crucial for obtaining information about the
mechanism or mechanisms that induce the decay.
In this section we investigate the possibility to discriminate between
different pairs of CP non-conserving mechanisms inducing the
 $\betabeta$-decay by using
data on $\betabeta$-decay half-lives of nuclei
with largely different NMEs.
 In addition to the nuclei  $^{76}$Ge, $^{82}$Se, $^{100}$Mo, $^{130}$Te
we will employ also the isotope $^{136}$Xe. Four sets of nuclear matrix
elements (NMEs) of the decays of these
five nuclei, derived within the
Self-consistent Renormalized
Quasiparticle Random Phase Approximation
(SRQRPA), will be employed in our analysis.
The analysis we are going to present is based on the fact that  for each of the five single
mechanisms discussed in \cite{Meroni:2012qf}, the NMEs for $^{76}$Ge, $^{82}$Se,
$^{100}$Mo and $^{130}$Te differ relatively little ---being
the relative difference between the NMEs of any two
nuclei not exceeding 10\%. The NMEs for $^{136}$Xe instead
differ significantly from those of
$^{76}$Ge, $^{82}Se$, $^{100}$Mo and $^{130}$Te,
being by a factor $\sim (1.3 - 2.5)$ smaller.
This allows, in principle, to draw conclusions about
the pair of non-interfering (interfering)
mechanisms possibly inducing the $\betabeta$-decay
from data on the half-lives of $^{136}Xe$ and
of at least one of the other used isotopes.
We will employ the lower bound obtained by the
EXO collaboration on the $\betabeta$-decay half-life of $^{136}$Xe
\cite{Auger:2012ar}:
\be T_{1/2}^{0\nu}(^{136}Xe) > 1.6 \times 10^{25} \rm{ y\quad \,
(90\,\% \,CL)}. \label{EXO1} \ee
Suppose now we analyse the case of two non-interfering mechanisms i.e.  that $T_i\equiv
T^{0\nu}_{1/2}(^{136}$Xe$)$, $T_j\equiv T^{0\nu}_{1/2}(^{76}$Ge$)$ and
that the \betabeta-decay is due by the standard light neutrino
exchange and the heavy RH Majorana neutrino exchange. In this case
the positivity conditions for $|\eta_\nu|^2$ and $|\eta_{_R}|^2$ imply
for the  Argonne [and CD-Bonn] NMEs corresponding to $g_A=1.25\,
(1.0)$:
\be 1.90\, (1.85) \left[1.30\, (1.16)  \right]\leq
\frac{T^{0\nu}_{1/2}(^{76}Ge)}{T^{0\nu}_{1/2}(^{136}Xe)} \leq
2.70\,(2.64), \left[2.47\,(2.30)\right]  \,; \label{nuRHNAr} \ee
%
Using the EXO result, eq. (\ref{EXO1}), and the Argonne NMEs we get
the lower bound on $ T^{0\nu}_{1/2}(^{76}$Ge$)$:
\be T^{0\nu}_{1/2}(^{76}Ge)\geq 3.03 \,(2.95)\times 10^{25}\,\rm{y}.
\label{Arg125} \ee
%
This lower bound is significantly bigger that the experimental lower
bound on $T^{0\nu}_{1/2}(^{76}$Ge$)$ quoted in eq. (\ref{limit}). If
we use instead the CD-Bonn NMEs,
the limit we obtain is close to the experimental lower bound on
$T^{0\nu}_{1/2}(^{76}$Ge$)$:
\be T^{0\nu}_{1/2}(^{76}Ge)\geq  2.08\,(1.85)\times 10^{25}\,\rm{y}.
\label{CDBonn125} \ee
%

\begin{figure}[h!]
   \begin{center}
 \subfigure
   {\includegraphics[width=6cm]{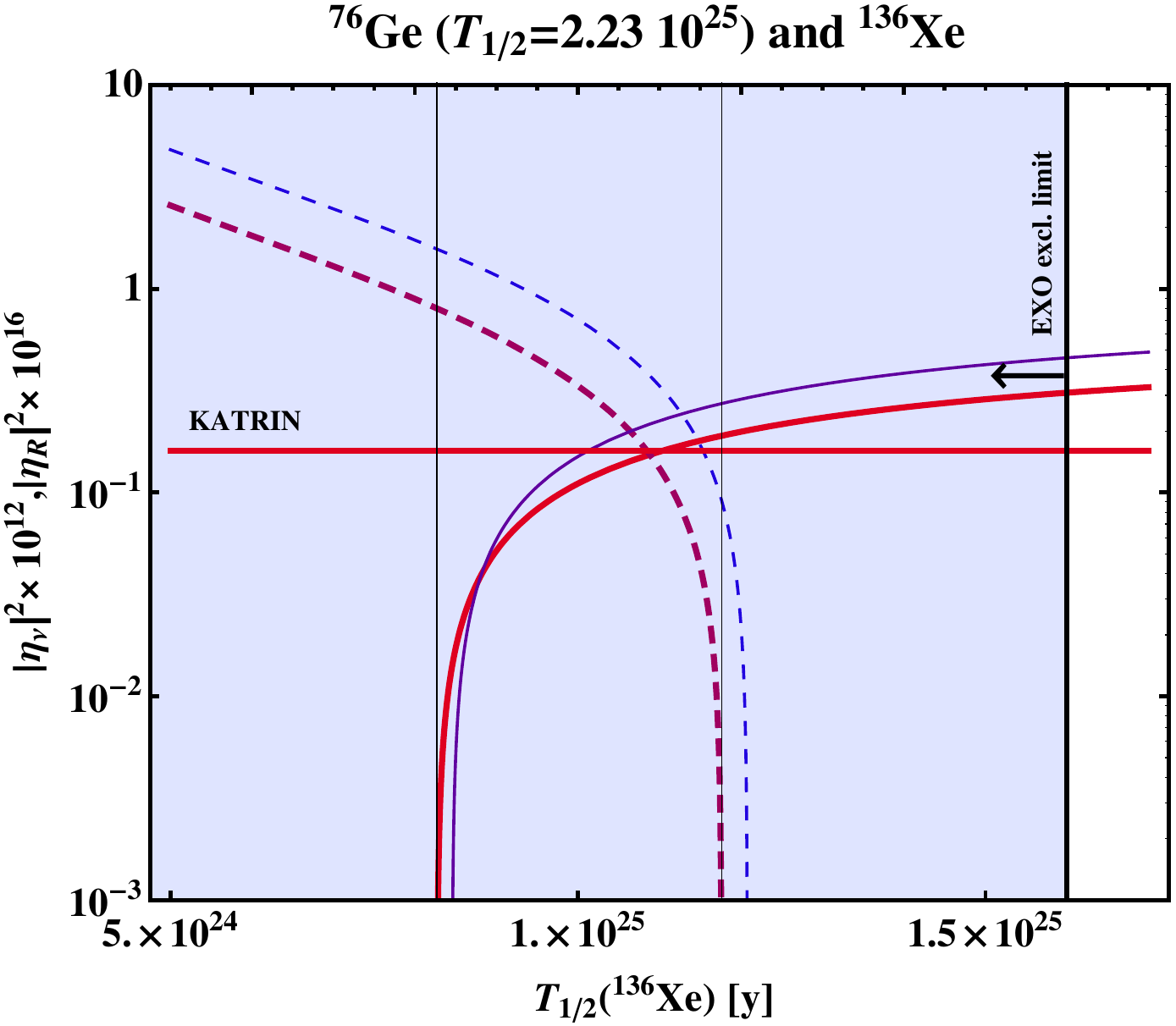}}
 \subfigure
   {\includegraphics[width=6cm]{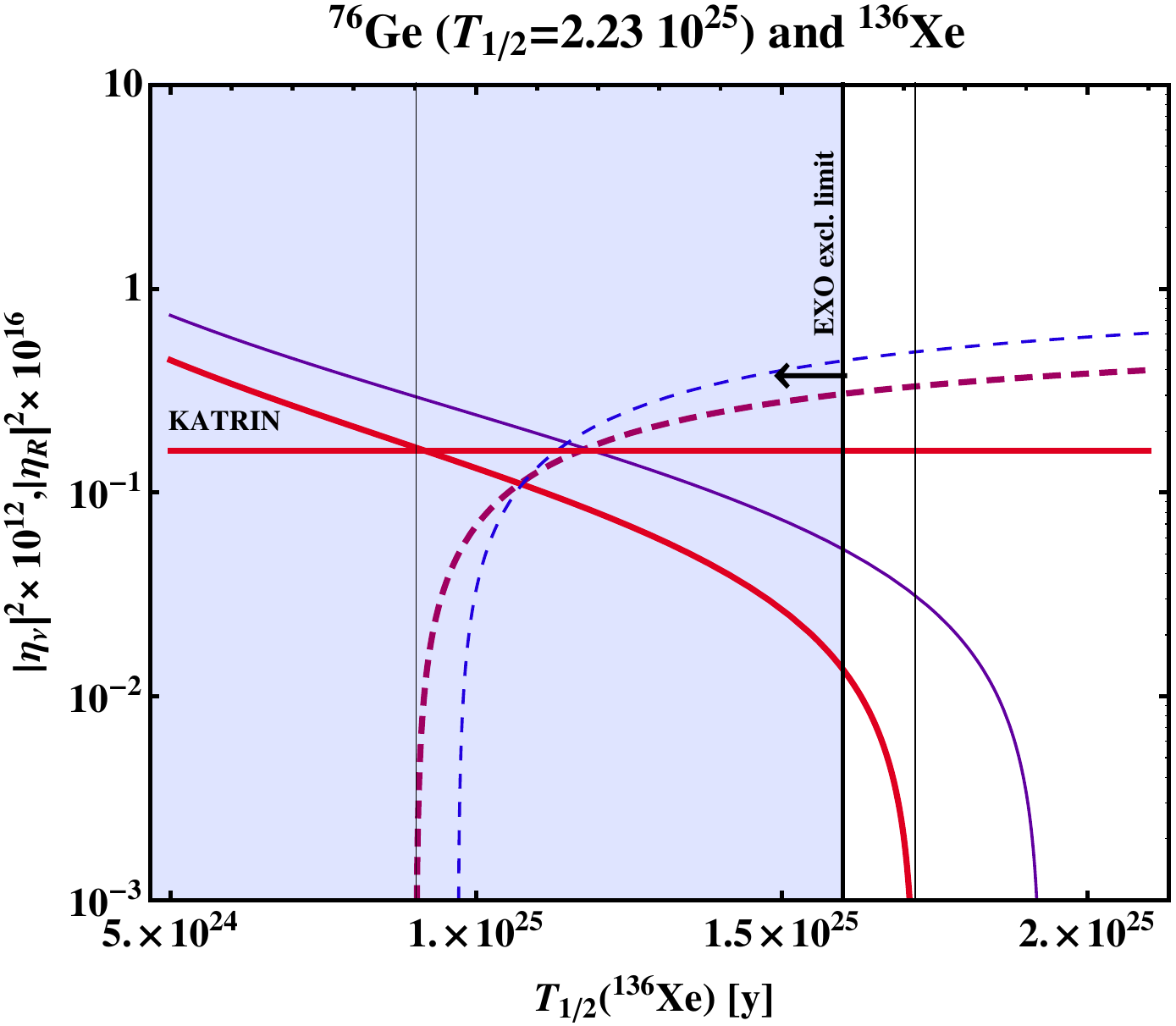}}
     \end{center}
 \vspace{-0.8cm}
 \caption{
\label{fig:figILL} The values of $|\eta_\nu|^2$ (solid lines) and
$|\eta_{_R}|^2$ (dashed lines) obtained for
$T^{0\nu}_{1/2}(^{76}$Ge$)=2.23 \times 10^{25}$ y
\cite{KlapdorK:2006ff} as a function of
$T^{0\nu}_{1/2}(^{136}$Xe$)$, using the Argonne (left panel) and
CD-Bonn (right panel) NMEs corresponding to $g_A=1.25$ (thick lines)
and  $g_A=1$ (thin lines). The region of physical (positive)
solutions for $g_A=1.25$ are delimited by the two vertical lines.
The solid horizontal line corresponds to the prospective upper limit
from the KATRIN experiment \cite{Eitel:2005hg}, while the thick solid
vertical line indicates the EXO lower bound  \cite{Auger:2012ar}.
The gray areas correspond to excluded values of $|\eta_\nu|^2$ and
$|\eta_{_R}|^2$. }
\end{figure}

For illustrative purposes we  show in Fig. \ref{fig:figILL} the
solutions of equation (\ref{solnonint}) for $|\eta_\nu|^2$ and
$|\eta_{_R}|^2$ derived by fixing  $T^{0\nu}_{1/2}(^{76}$Ge$)$ to the
best fit value claimed in \cite{KlapdorK:2006ff},
$T^{0\nu}_{1/2}(^{76}$Ge$)=2.23 \times 10^{25}$ (see eq.
(\ref{limit})). As Fig. \ref{fig:figILL} shows, the positive
(physical) solutions obtained using the Argonne NMEs are
incompatible with the EXO result, eq. (\ref{EXO1}), and under the
assumptions made and according to our oversimplified analysis, are
ruled out. At the same time, the physical solutions obtained using
the CD-Bonn NMEs are compatible with the EXO limit for values of
$|\eta_\nu|^2$ and $|\eta_{_R}|^2$ lying in a relatively narrow
interval.

As
we noticed in \cite{Faessler:2011qw}, if the experimentally determined
interval of allowed values of the ratio $T_j/T_i$ of the half-lives
of the two isotopes considered, including all relevant
uncertainties, lies outside the range of positive solutions for
$|\eta_A|^2$ and $|\eta_B|^2$, one would be led to conclude that the
$\betabeta$-decay is not generated by the two mechanisms under
discussion.

%
\section{Discriminating between Different
Pairs of  Non-interfering Mechanisms}
%

\begin{figure}[htbp]
\centering
\begin{center}
 \subfigure
   { \includegraphics[width=0.4 \textwidth]{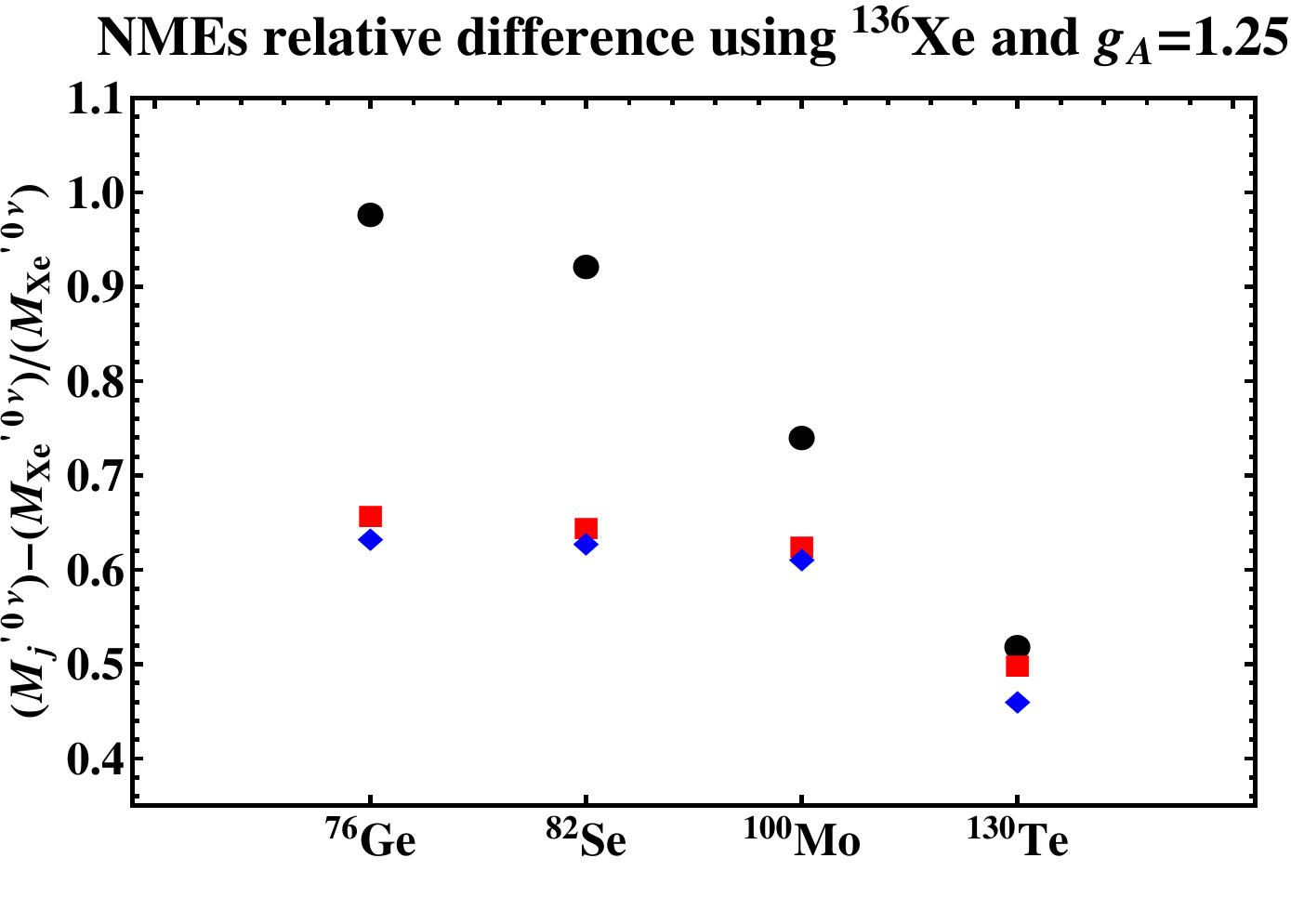}}
 \subfigure
   {\includegraphics[width=0.4 \textwidth]{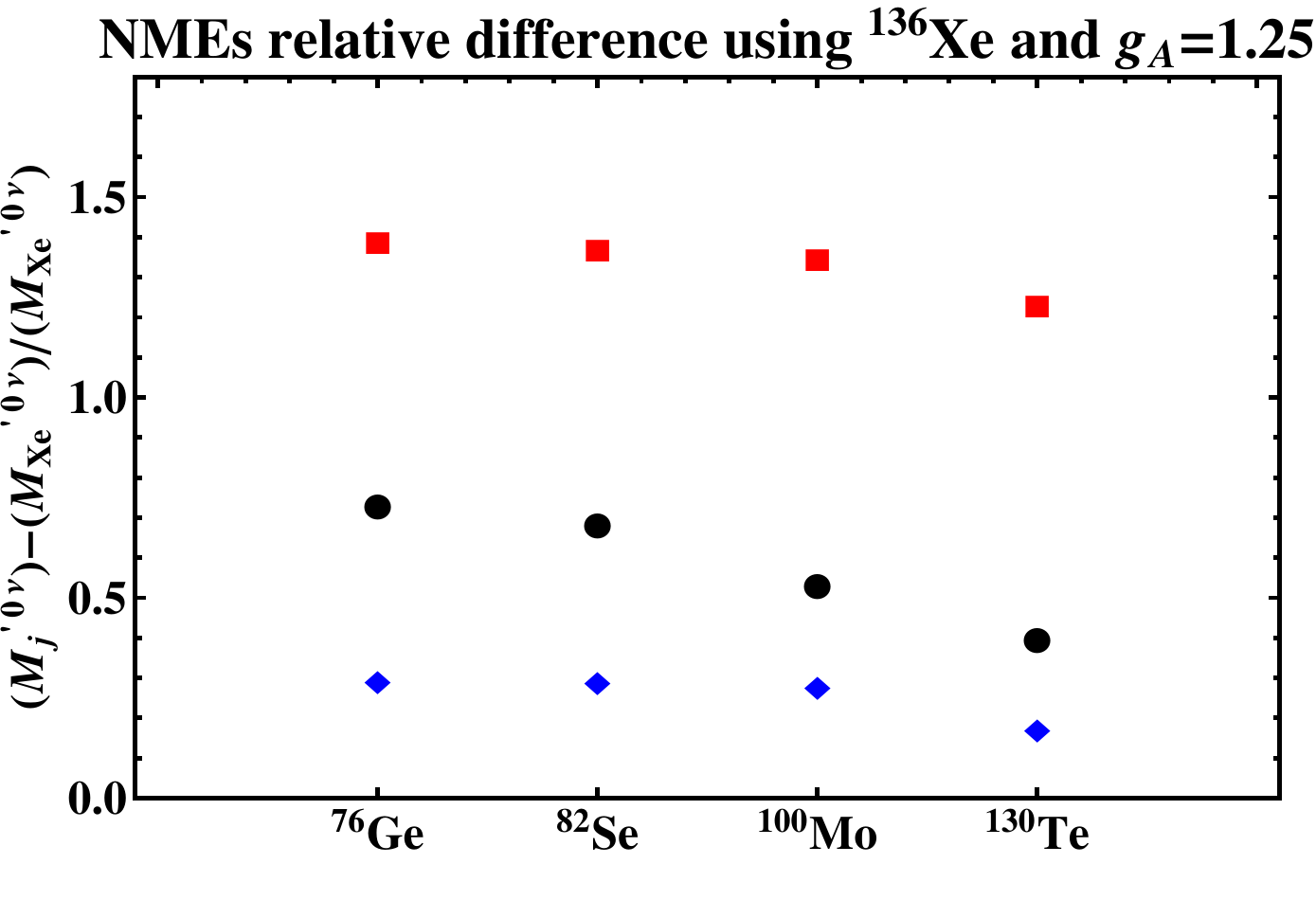}}
   \end{center}
 \vspace{-0.8cm}
   \caption{\label{fig:diff1} The relative differences
between the Argonne  and Cd-Bonn NMEs
$({\cal M}^{0\nu}_{j}-{\cal M}^{0\nu}_{i})/{\cal M}^{0\nu}_{i}$, where
$i$=$^{136}$Xe and $j$ =$^{76}$Ge,$^{82}$Se,$^{100}$Mo,$^{130}$Te,
for $g_A=1.25$  and for three
different non-interfering mechanisms: light Majorana neutrino
exchange (circles), RH heavy Majorana neutrino exchange (squares)
and gluino exchange (diamonds). See text for details.}
\end{figure}

The first thing to notice is that for
each of the four different mechanisms of $\betabeta$-decay
considered, the relative difference between NMEs of the decays of
$^{76}$Ge, $^{82}$Se, $^{100}$Mo and $^{130}$Te does not exceed
approximately 10\%: $({\cal M}^{0\nu}_{j,X} -
{\cal M}^{0\nu}_{i,X})/(0.5({\cal M}^{0\nu}_{j,X} + {\cal M}^{0\nu}_{i,X}))
\ltap 0.1$, where $i\neq j$ =
$^{76}$Ge,$^{82}$Se,$^{100}$Mo,$^{130}$Te, and $X$ denotes any one
of the four mechanisms discussed. As was shown in the previous section
this leads to degeneracies between the positivity intervals of
values of the ratio of the half-lives of any two given of the
indicated four isotopes, corresponding to the different pairs of
mechanisms inducing the $\betabeta$-decay. The degeneracies in
question make it practically impossible to distinguish between the
different pairs of $\betabeta$-decay mechanisms, considered in
the previous section and in the present section, using data on the
half-lives of two or more of the four nuclei $^{76}$Ge, $^{82}$Se,
$^{100}$Mo and $^{130}$Te. At the same time, it is possible, in
principle, to exclude them all using data on the half-lives of at
least two of the indicated four nuclei.
 In contrast, the NMEs for the $\betabeta$-decay of
$^{136}Xe$, corresponding to each of the four different mechanisms
we are considering are by a factor of $\sim (1.3 - 2.5)$ smaller
than the $\betabeta$-decay NMEs of the other four isotopes listed
above: $({\cal M}^{0\nu}_{j,X} - {\cal M}^{0\nu}_{i,X})/{\cal M}^{0\nu}_{i,X})
\cong (0.3-1.5)$, where $i$ = $^{136}Xe$ and $j$ =
$^{76}$Ge,$^{82}$Se,$^{100}$Mo,$^{130}$Te (see Figs. \ref{fig:diff1}). As a consequence, using data on the half-life
of $^{136}Xe$ as input in determining the positivity interval of
values of the half-life of any second isotope lifts to a certain
degree the degeneracy of the positivity intervals corresponding to
different pairs of non-interfering mechanisms. This allows, in
principle, to draw conclusions about the pair of mechanisms possibly
inducing the $\betabeta$-decay from data on the half-lives of
$^{136}Xe$ and a second isotope which can be, e.g., any of the four
considered above,  $^{76}$Ge, $^{82}$Se, $^{100}$Mo and $^{130}$Te.

Therefore, we analyze next the possibility to
discriminate between two pairs of non-interfering mechanisms
triggering the $\betabeta$-decay when the pairs share one mechanism.
Given three different non-interfering mechanisms $A$, $B$ and $C$,
we can test the hypothesis of the $\betabeta$-decay induced by the
pairs i) $A+B$ or  ii) $C+B$, using the half-lives of the same two
isotopes. As  a consequence of the fact that B is common to both
pairs of mechanisms, the numerators of the expressions for
$|\eta_A|^2$ and $|\eta_C|^2$, as it follows from eq.
(\ref{solnonint}), coincide. Correspondingly, using the half-lives
of the same two isotopes would allow us to distinguish, in
principle, between the cases i) and ii) if the denominators in the
expressions for the solutions for $|\eta_A|^2$ and  $|\eta_C|^2$
have opposite signs. Indeed, in this case the physical solutions for
$|\eta_A|^2$ in the case i) and $|\eta_C|^2$ in the case ii)  will
lie either in the positivity intervals, see e.g. eq. \eqref{PosC}. Thus, the positivity solution intervals for
$|\eta_A|^2$ and $|\eta_C|^2$ would not overlap, except for the
point corresponding to a value of the second isotope half-life where
$\eta_A = \eta_C = 0$. This would allow, in principle, to
discriminate between the two considered pairs of mechanisms.

 It follows from the preceding discussion that
in order to be possible to discriminate between the pairs $A+B$ and
$C+B$ of non-interfering mechanisms of $\betabeta$-decay, the
following condition has to be fulfilled:
\be \frac{Det \begin{pmatrix}
|{\cal M}^{0\nu}_{i,A}|^2 & |{\cal M}^{0\nu}_{i,B}|^2 \\
 |{\cal M}^{0\nu}_{j,A }|^2& |{\cal M}^{0\nu}_{j,B
}|^2\end{pmatrix}  }{   Det \begin{pmatrix}
|{\cal M}^{0\nu}_{i,C}|^2 & |{\cal M}^{0\nu}_{i,B}|^2 \\
 |{\cal M}^{0\nu}_{j,C }|^2& |{\cal M}^{0\nu}_{j,B
}|^2\end{pmatrix}    }=\frac{|{\cal M}^{0\nu}_{i,A}|^2|{\cal M}^{0\nu}_{j,B
}|^2-|{\cal M}^{0\nu}_{i,B}|^2 |{\cal M}^{0\nu}_{j,A
}|^2}{|{\cal M}^{0\nu}_{i,C}|^2|{\cal M}^{0\nu}_{j,B
}|^2-|{\cal M}^{0\nu}_{i,B}|^2 |{\cal M}^{0\nu}_{j,C }|^2}<0\,. \ee
%
This condition is satisfied if one of the following two sets of
inequalities holds:
\ba \label{ineqI} I)&&
\frac{{\cal M}^{0\nu}_{j,C}-{\cal M}^{0\nu}_{i,C}}{{\cal M}^{0\nu}_{i,C}} <
\frac{{\cal M}^{0\nu}_{j,B}-{\cal M}^{0\nu}_{i,B}}{{\cal M}^{0\nu}_{i,B}} <
\frac{{\cal M}^{0\nu}_{j,A}-{\cal M}^{0\nu}_{i,A}}{{\cal M}^{0\nu}_{i,A  }},\\
\label{ineqII} II)&&
\frac{{\cal M}^{0\nu}_{j,A}-{\cal M}^{0\nu}_{i,A}}{{\cal M}^{0\nu}_{i,A }}    <
\frac{{\cal M}^{0\nu}_{j,B}-{\cal M}^{0\nu}_{i,B}}{{\cal M}^{0\nu}_{i,B}} <
\frac{{\cal M}^{0\nu}_{j,C}-{\cal M}^{0\nu}_{i,C}}{{\cal M}^{0\nu}_{i,C}}.
 \ea
%
One example of a possible application of the preceding results is
provided by the mechanisms of light Majorana neutrino exchange (A),
RH heavy Majorana neutrino exchange (B) and gluino exchange (C) and
the Argonne NMEs. We are interested in studying cases involving
$^{136}$Xe since, as it was already discussed  earlier, the NMEs of
$^{136}$Xe differ significantly from those of the lighter isotopes
such as $^{76}$Ge. Indeed, as can be
shown, it is possible, in principle, to discriminate between the two
pairs $A+B$ and $B+C$ of the three mechanisms indicated above if we
combine data on the half-life of $^{136}$Xe with those on the
half-life of one of the four isotopes $^{76}$Ge, $^{82}$Se,
$^{100}$Mo and $^{130}$Te, and use the Argonne NMEs in the analysis.
In this case the inequalities (\ref{ineqI}) are realized, as can be
seen in the left panel of Fig. \ref{fig:diff1}, where we plot the relative differences
$({\cal M}^{0\nu}_{j}-{\cal M}^{0\nu}_{i})/{\cal M}^{0\nu}_{i}$ for the Argonne
NMEs where the indices $i$ and $j$ refer respectively to $^{136}Xe$
and to one of the four isotopes $^{76}$Ge, $^{82}$Se, $^{100}$Mo and
$^{130}$Te. In the case of the CD-Bonn NMEs (right panel of Fig. \ref{fig:diff1}),
the inequalities (\ref{ineqI}) or (\ref{ineqII}) do not hold for the
pairs of mechanisms considered. The inequalities given in eq.
(\ref{ineqI}) hold, as it follows from right panel of Fig. \ref{fig:diff1}, if,
e.g., the mechanisms A, B and C are respectively the heavy RH
Majorana neutrino exchange, the light Majorana neutrino exchange and
the gluino exchange (for a definition  of the gluino exchange mechanism see \cite{Faessler:2011qw, Meroni:2012qf}, and references therein). In the following we will denote the latter with $\eta_{\lambda'}$.

\begin{figure}[htbp]
\centering%
\subfigure
   {\includegraphics[width=6cm]{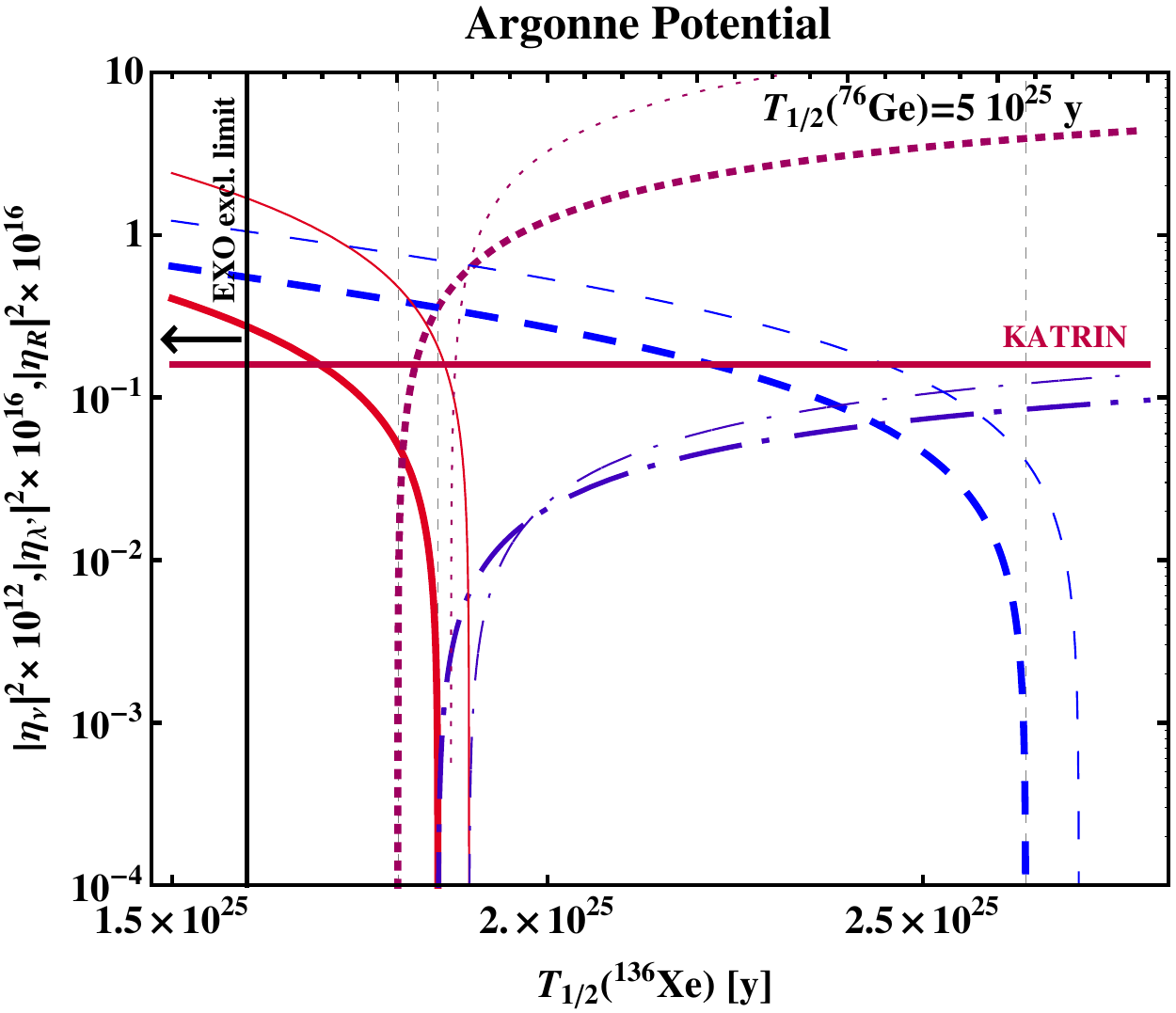}}
 \vspace{2mm}
 \subfigure
   {\includegraphics[width=6cm]{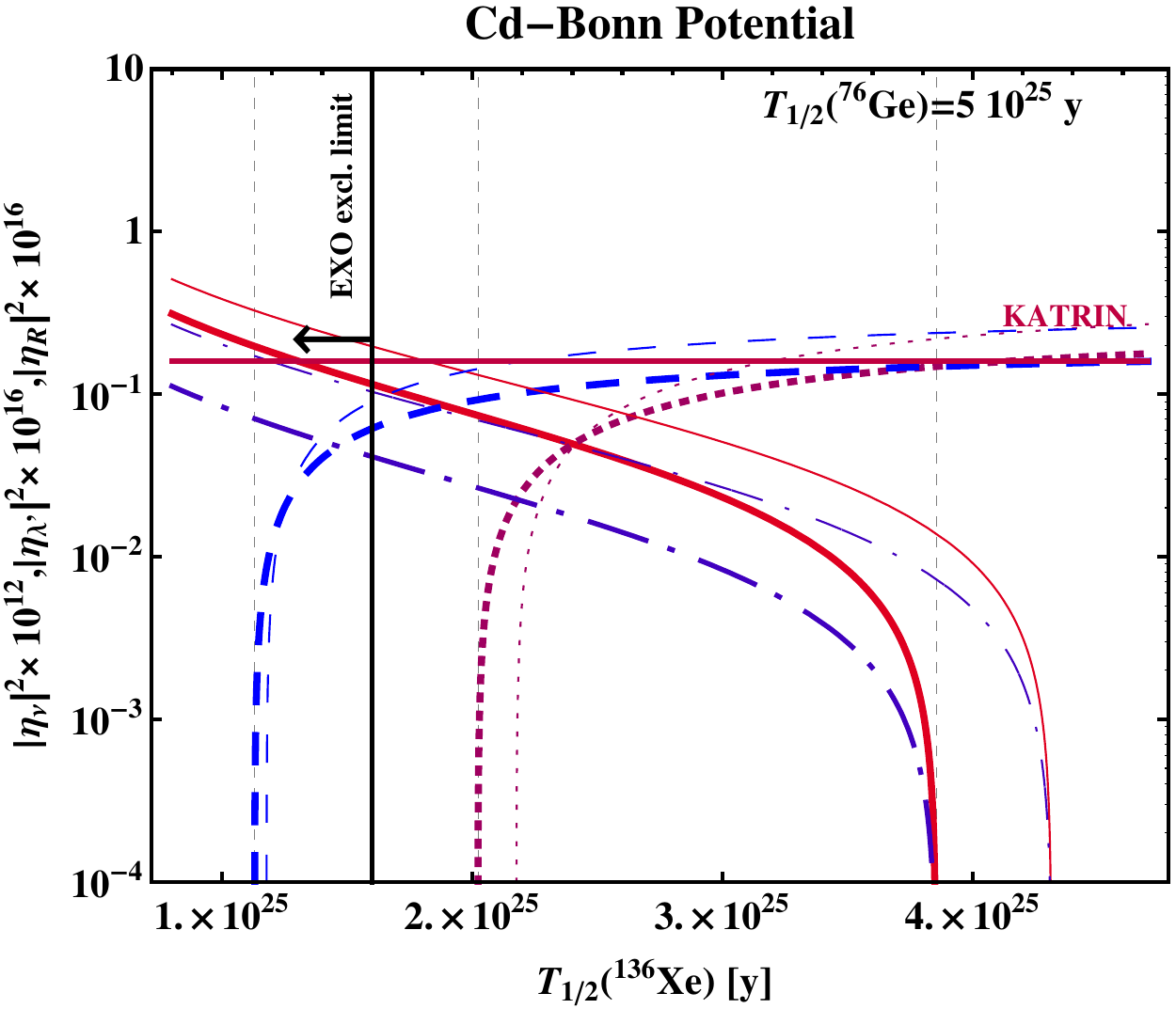}}
 \caption{\label{fig:comparison}
Solutions for the LNV parameters corresponding to two pairs of
non-interfering mechanisms: i) $|\eta_\nu|^2$ and $|\eta_{_R}|^2$
(dot-dashed  and dashed lines)  and ii) $|\eta_{\lambda'}|^2$ and
$|\eta_{_R}|^2$ (solid and dotted lines). The solutions are obtained
by fixing $T_i = T^{0\nu}_{1/2}(^{76}$Ge$)=5 \times 10^{25}$ y and
letting free $T_j = T^{0\nu}_{1/2}(^{136}$Xe$)$ and using the sets of
Argonne (left panel) and CD-Bonn (right panel) NMEs  calculated for
$g_A= 1.25$ (thick lines) and $g_A=1$ (thin lines). The range of
positive solutions in the case of Argonne NMEs and $g_A=1.25$ is
delimited by the two vertical dashed lines. 
}
\end{figure}

\begin{figure}[htbp]
\centering%
\subfigure
   {\includegraphics[width=6.7cm]{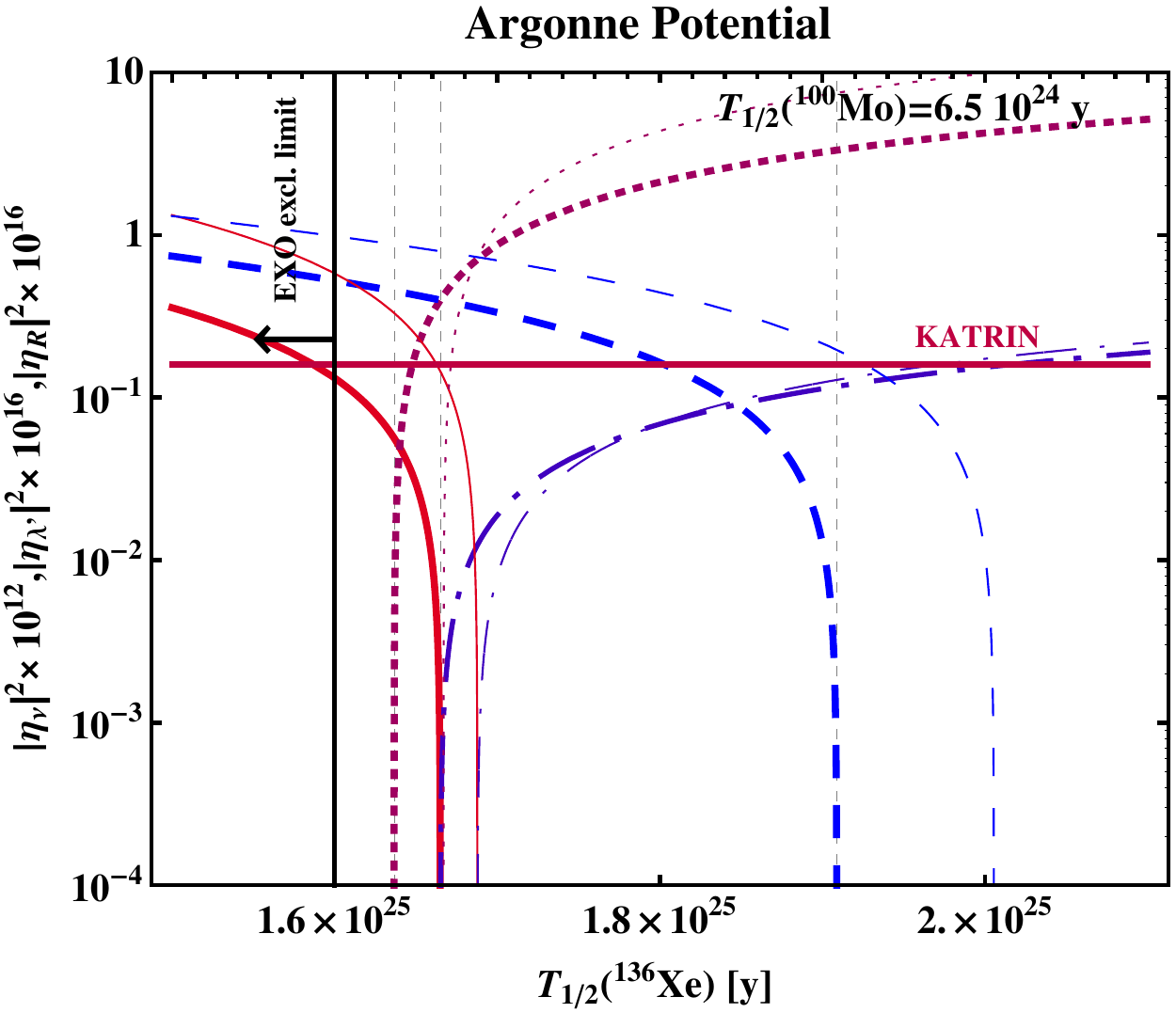}}
 \vspace{2mm}
 \subfigure
   {\includegraphics[width=7cm]{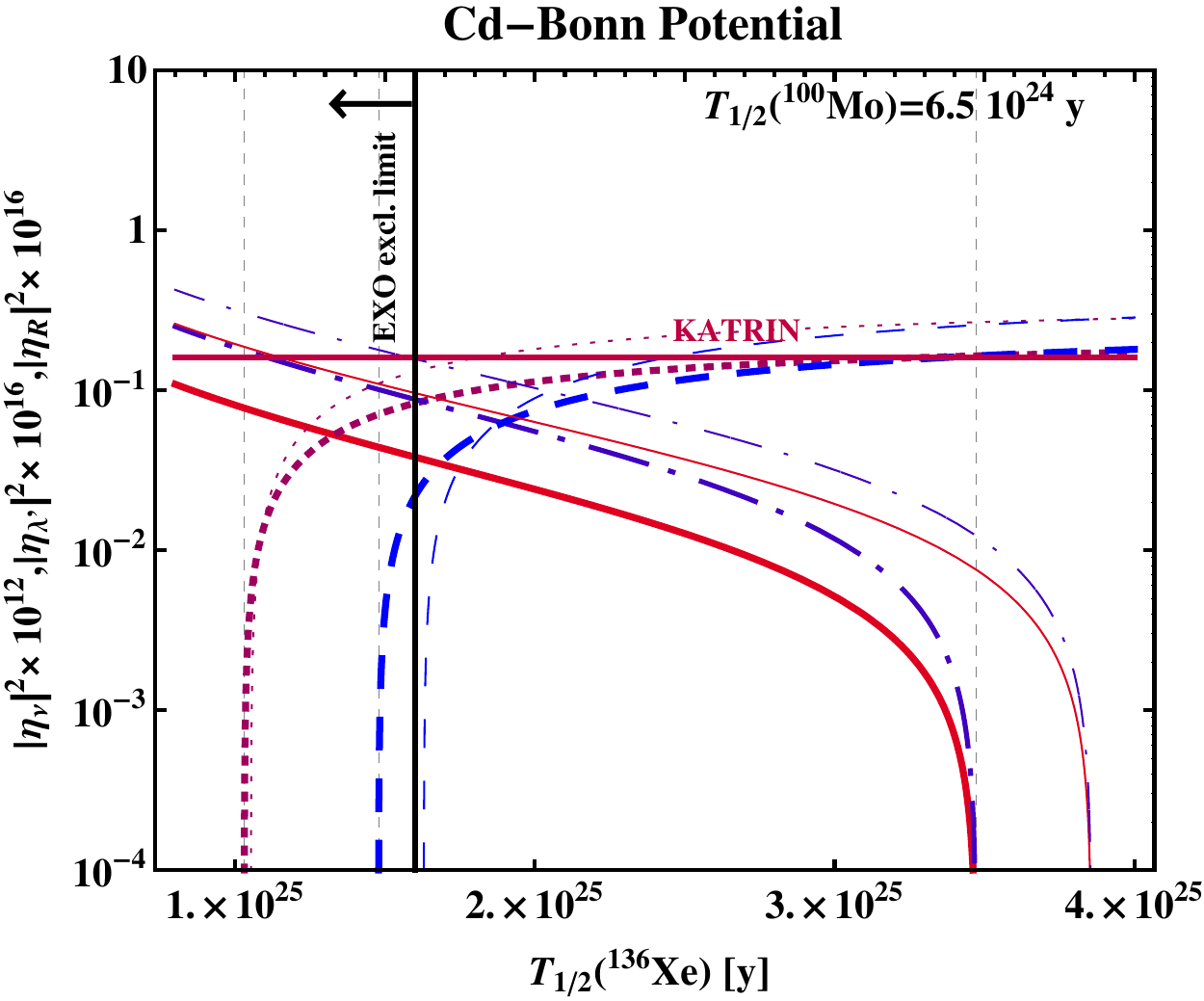}}
 \caption{\label{fig:comparison2}
Solutions for the LNV parameters of  two pairs of non-interfering
$\betabeta$-decay mechanisms i) $|\eta_\nu|^2$ and $|\eta_{_R}|^2$
(dot-dashed  and dashed lines)  and ii) $|\eta_{\lambda'}|^2$ and
$|\eta_{_R}|^2$ (solid and dotted lines) obtained by fixing $T_i =
T^{0\nu}_{1/2}(^{100}$Mo$)=6.5 \times 10^{24}$ yr and letting free
$T_j = T^{0\nu}_{1/2}(^{136}$Xe$)$. 
 }
\end{figure}

 The preceding considerations are illustrated graphically
in Figs. \ref{fig:comparison} and \ref{fig:comparison2}. In  Fig.
\ref{fig:comparison} we use $T_i \equiv T^{0\nu}_{1/2}(^{76}$Ge$)$
and  $T_j \equiv T^{0\nu}_{1/2}(^{136}$Xe$)$ and the Argonne (left
panel) and CD-Bonn (right panel) NMEs for the decays of $^{76}$Ge
and $^{136}$Xe to show the possibility of discriminating between the
two pairs of non-interfering mechanisms considered earlier: i) light
Majorana neutrino exchange and heavy RH Majorana neutrino exchange
(RHN) and ii) heavy RH Majorana neutrino exchange and gluino
exchange. The $^{76}$Ge half-life is set to $T_i = 5 \times 10^{25}$
y, while that of  $^{136}$Xe, $T_j$, is allowed to vary in a certain
interval. The solutions for the three LNV parameters corresponding
to the three mechanisms considered, $|\eta_\nu|^2$, $|\eta_{_R}|^2$ and
$|\eta_{\lambda'}|^2$, obtained for the chosen value of $T_i$ and
interval of values of $T_j$, are shown as functions of $T_j$. As is
clearly seen in the left panel of Fig. \ref{fig:comparison}, if
$|\eta_\nu|^2$, $|\eta_{_R}|^2$ and $|\eta_{\lambda'}|^2$ are obtained
using the Argonne NMEs, the intervals of values of $T_j$ for which
one obtains the physical positive solutions for $|\eta_\nu|^2$ and
$|\eta_{\lambda'}|^2$, do not overlap. This  makes it possible, in
principle, to determine which of the two pairs of mechanisms
considered (if any) is inducing  the $\betabeta$-decay. The same
result does not hold if one uses the CD-Bonn NMEs in the analysis,
as is illustrated in the right panel of Fig. \ref{fig:comparison}.
In this case none of the inequalities (\ref{ineqI}) and
(\ref{ineqII}) is fulfilled, the intervals of values of $T_j$ for
which one obtains physical solutions for $|\eta_\nu|^2$ and
$|\eta_{\lambda'}|^2$ overlap and the discrimination between the two
pairs of mechanisms is problematic.

 We show in Fig. \ref{fig:comparison2} that the
features of the solutions for $|\eta_\nu|^2$ and
$|\eta_{\lambda'}|^2$ we have discussed above, which are related to
the values of the relevant NMEs, do not change if one uses in the
analysis the half-lives and NMEs of $^{136}Xe$ and of another
lighter isotope instead of $^{76}$Ge, namely, of $^{100}$Mo.
%
%
\section{Two Interfering Mechanisms}
%

  We analyze in the present Section the possibility of
$\betabeta$-decay induced by two interfering CP-non-conserving
mechanisms. As we have seen in the previous section this case is characterized by three parameters: the
absolute values and the relative phase of the two LNV parameters
associated with the two mechanisms. They can be determined, in
principle, from data on the half-lives of three isotopes, $T_i$,
$i=1,2,3$. Given $T_{1,2,3}$ and denoting by $A$ and $B$ the two
mechanisms, one can set a system of three linear equations in three
unknowns, the solution of which reads:
\be |\eta_A|^2= \frac{D_i}{D}\,,~~ |\eta_{B}|^2 = \frac{D_j}{D}\,,~~
z \equiv 2\cos\alpha|\eta_A||\eta_{B}| = \frac{D_k}{D}\,,
\label{intsol1} \ee
%
where $D$, $D_i$, $D_j$ and $D_k$ are the following determinants:
\be D = \left| \begin{array}{ccc}
 ({\cal M}^{0\nu}_{i,A })^2 & ({\cal M}^{0\nu }_{i,B })^2 & {\cal M}^{0\nu }_{i,B } {\cal M}^{0\nu}_{i,A } \\
 ({\cal M}^{0\nu}_{j,A })^2 & ({\cal M}^{0\nu }_{j,B })^2 & {\cal M}^{0\nu }_{j,B } {\cal M}^{0\nu}_{j,A }\\
 ({\cal M}^{0\nu}_{k,A })^2 & ({\cal M}^{0\nu }_{k,B})^2 & {\cal M}^{0\nu }_{k,B } {\cal M}^{0\nu}_{k,A }
\end{array}
\right|\,,~~~ D_i = \left| \begin{array}{ccc}
i/ T_i G_i & ({\cal M}^{0\nu }_{i,B})^2 & {\cal M}^{0\nu }_{i,B} {\cal M}^{0\nu}_{i,A } \\
i/ T_j G_j & ({\cal M}^{0\nu }_{j,B})^2 & {\cal M}^{0\nu }_{j,B} {\cal M}^{0\nu}_{j,A } \\
i/ T_k G_k & ({\cal M}^{0\nu }_{k,B})^2 & {\cal M}^{0\nu}_{k,B}
{\cal M}^{0\nu}_{k,A }
\end{array} \right|\,,
\label{DDi} \ee
\be D_j=  \left| \begin{array}{ccc}
({\cal M}^{0\nu}_{i,A})^2 & i/ T_i G_i &  {\cal M}^{0\nu }_{i,B} {\cal M}^{0\nu}_{i,A } \\
({\cal M}^{0\nu}_{j,A })^2 & i/ T_j G_j &  {\cal M}^{0\nu }_{j,B} {\cal M}^{0\nu}_{j,A } \\
({\cal M}^{0\nu}_{k,A })^2 & i/ T_k G_k &  {\cal M}^{0\nu }_{k,B}
{\cal M}^{0\nu}_{k,A }
\end{array}
\right|\,,~~~ D_k = \left| \begin{array}{ccc}
  ({\cal M}^{0\nu}_{i,A })^2& ({\cal M}^{0\nu }_{i,B})^2 & i/ T_i G_i \\
  ({\cal M}^{0\nu}_{j,A })^2& ({\cal M}^{0\nu }_{j,B})^2 & i/ T_j G_j\\
  ({\cal M}^{0\nu}_{k,A })^2& ({\cal M}^{0\nu }_{k,B})^2 & i/ T_k G_k
\end{array} \right|\,.
\label{D2Dk} \ee
%
As in the case of two non-interfering mechanisms, the LNV parameters
must be non-negative $|\eta_A|^2 \geq  0$ and $|\eta_{B}|^2 \geq
0$, and in addition the interference term must satisfy the following
condition:
\be -2|\eta_A||\eta_{B}| \leq 2\cos\alpha|\eta_A||\eta_{B}|
\leq2|\eta_A||\eta_{B}|\,. \label{fase} \ee
%
These conditions will be called from here on ``positivity
conditions''.

  Using the positivity conditions it is possible
to  determine the interval of positive solutions for one of the
three half-life, e.g., $T_k$, if the values of the other two
half-lives in the equations have been measured and are known. The
condition on the interference term in equation (\ref{PosC}) can
considerably reduce the interval of values of $T_k$ where
$|\eta_{A}|^2\geq 0$ and $|\eta_{B}|^2\geq 0$. In Table
\ref{tab:tabint} we give examples of the constraints on $T_k$
following from the positivity conditions for three different pairs
of interfering mechanisms: light Majorana neutrino and
supersymmetric gluino exchange; light Majorana neutrino exchange and
heavy LH Majorana neutrino exchange; gluino exchange and heavy LH
Majorana neutrino exchange. It follows from the results shown in
Table \ref{tab:tabint}, in particular, that when $T^{0\nu}_{1/2}(^{76}$Ge$)$ is set
to  $T^{0\nu}_{1/2}(^{76}$Ge$)=2.23\times 10^{25};10^{26}$ y, but  $T^{0\nu}_{1/2}(^{130}$Te$)$ is
close to the current experimental lower limit, the positivity
constraint intervals of values of   $T^{0\nu}_{1/2}(^{136}$Xe$)$ for the each of the
three pairs of interfering mechanisms considered are incompatible
with the EXO lower bound on   $T^{0\nu}_{1/2}(^{136}$Xe$)$, eq. (\ref{EXO1}).
\begin{table}[h!]
\centering \caption{ \label{tab:tabint} Ranges of the half-live of
$^{136}$Xe for different fixed values of the half-lives of $^{76}$Ge
and $^{130}$Te in the case of three  pairs of interfering
mechanisms: light Majorana neutrino exchange and gluino exchange
(upper table); light Majorana and heavy LH Majorana neutrino
exchanges (lower table). The results shown are obtained with
the ``large basis''  $g_A=1.25$  Argonne NMEs.  Two stars
indicates that the EXO bound  rules out the
corresponding solution. } \vspace{15pt}
\begin{tabular}{|l|l|c|}
\toprule
 T$^{0\nu}_{1/2}$[y](fixed) &  T$^{0\nu}_{1/2}$[y](fixed) & Allowed Range \\
\midrule
  T$(Ge)=2.23\cdot10^{25}$**  &  T$(Te)=3\cdot10^{24}$     &   $ 2.95 \cdot 10^{24}\leq T(Xe) \leq 5.65\cdot 10^{24}$ \\
          T$(Ge)= 10^{26}$**  &  T$(Te)=3\cdot10^{24}$     & $ 3.43\cdot10^{24} \leq T(Xe) \leq   4.66\cdot10^{24}$\\
 T$(Ge)= 2.23\cdot10^{25}$  &  T$(Te)=3\cdot10^{25}$     &$ 1.74\cdot10^{25} \leq T(Xe) \leq1.66 \cdot10^{26}$\\
   T$(Ge)= 10^{26}$         &  T$(Te)=3\cdot10^{25}$     & $2.58 \cdot10^{25} \leq T (Xe) \leq 6.90 \cdot10^{25}$\\
\bottomrule
\end{tabular}\vspace{15pt}
\begin{tabular}{|l|l|c|}
\toprule
 T$^{0\nu}_{1/2}$[y](fixed) &  T$^{0\nu}_{1/2}$[y](fixed) & Allowed Range \\
\midrule
  T$(Ge)=2.23\cdot10^{25}$**  &  T$(Te)=3\cdot10^{24}$     &   $ 4.93\cdot 10^{24}\leq T(Xe) \leq6.21\cdot 10^{24}$ \\
          T$(Ge)= 10^{26}$** &  T$(Te)=3\cdot10^{24}$     & $5.23 \cdot10^{24} \leq T(Xe) \leq   5.83 \cdot10^{24}$\\
 T$(Ge)= 2.23\cdot10^{25}$  &  T$(Te)=3\cdot10^{25}$     &$ 3.95 \cdot10^{25} \leq T(Xe) \leq 8.25\cdot10^{25}$\\
   T$(Ge)= 10^{26}$         &  T$(Te)=3\cdot10^{25}$     & $4.68 \cdot10^{25} \leq T (Xe) \leq 6.61\cdot10^{25}$\\
\bottomrule
\end{tabular}
\end{table}
%

 We consider next
a case in which the half-life of $^{136}$Xe is one of the two
half-lives assumed to have been experimentally determined. The
\betabeta-decay is supposed to be triggered by light Majorana
neutrino and gluino exchange mechanisms with LFV parameters
$|\eta_\nu|^2$ and $|\eta_{\lambda'}|^2$. We use in the analysis the
half-lives of $^{76}$Ge, $^{136}$Xe and  $^{130}$Te, which will be
denoted for simplicity respectively as $T_1$, $T_2$ and $T_3$. Once
the experimental bounds on $T_i$,  $i=1,2,3$, given in eq.
(\ref{limit}), are taken into account, the conditions for
destructive interference, i.e., for $\cos\alpha <0$, are given by:
\be
 z < 0\,:
\begin{cases}
1.9\times 10^{25} < T_1 \leq 1.90 T_2,  &   T_3 \geq\dfrac{9.64 T_1 T_2}{16.32  T_1 +8.59 T_2}; \\
 1.90 T_2 < T_1 \leq 2.78 T_2, & T_3 >\dfrac{3.82 T_1 T_2}{6.33 T_1 + 3.66 T_2}; \\
 T_1 > 2.78 T_2, & T_3 \geq\dfrac{7.33 T_1 T_2}{11.94 T_1 +7.61 T_2}\,, \\
\end{cases}
\label{destrint} \ee
%
where we have used the ``large basis''  $g_A=1.25$ Argonne NMEs. The conditions for constructive interference
read:
\be
 z > 0\,:
\begin{cases}
1.90  T_2< T_1 \leq 2.29 T_2,  &  \dfrac{9.64 T_1 T_2}{16.32  T_1 +8.59 T_2} \leq T_3 \leq  \dfrac{3.82 T_1 T_2}{6.33 T_1 + 3.66 T_2} ;\\
 2.29 T_2< T_1 < 2.78 T_2, & \dfrac{7.33 T_1 T_2}{11.94 T_1 +7.61 T_2} \leq  T_3 \leq \dfrac{3.82 T_1 T_2}{6.33 T_1 + 3.66 T_2}.\\
\end{cases}
\label{constrint} \ee
%
If we set, e.g.,  the $^{76}$Ge half-life to the value claimed in
\cite{KlapdorK:2006ff} $T_1=2.23\times 10^{25}$ y, we
find that only destructive interference between the contributions of
the two mechanisms considered in the $\betabeta$-decay rate, is
possible. Numerically we get in this case
\be T_3 > \dfrac{3.44 T_2}{5.82  + 1.37 \times 10^{-25} T_2}\,.
\label{destrint2} \ee
%
For  $1.37 \times 10^{-25} T_2 \ll 5.82$ one finds:
\be T(^{130}Te) \gtrsim 0.59\, T(^{136}Xe) \gtrsim 9.46\times
10^{24}~\rm{y}\,, \label{destrint2v} \ee
%
where the last inequality has been obtained using the  EXO lower
bound on  $T(^{136}Xe)$. Constructive interference is possible for
the pair of interfering mechanisms under discussion only if
$T(^{76}Ge) \gtrsim 3.033 \times 10^{25}$ y.

\begin{figure}[h!]
  \begin{center}
 \subfigure
 {\includegraphics[width=6.9cm]{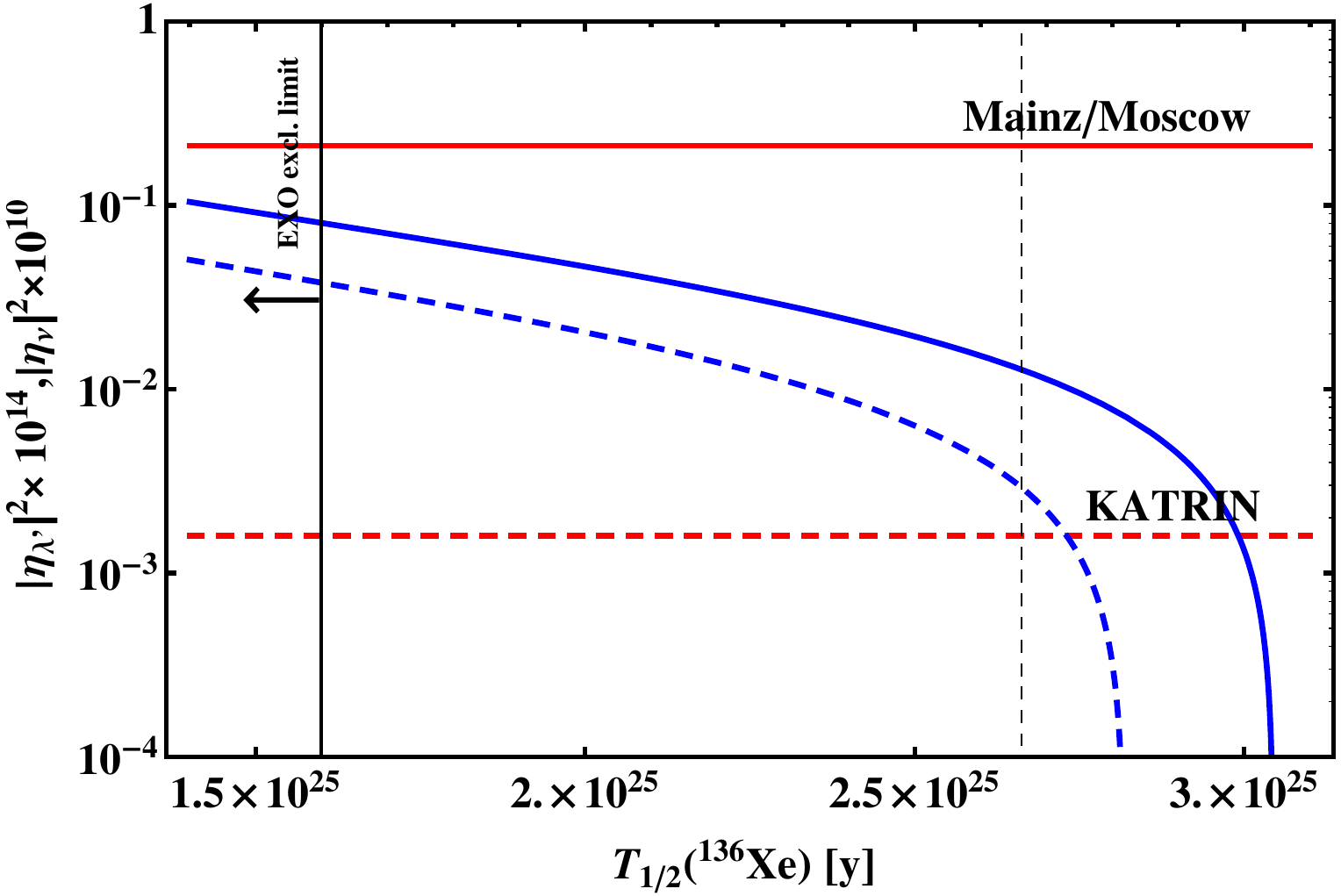}}
 \vspace{5mm}
 \subfigure
   {\includegraphics[width=6.6cm]{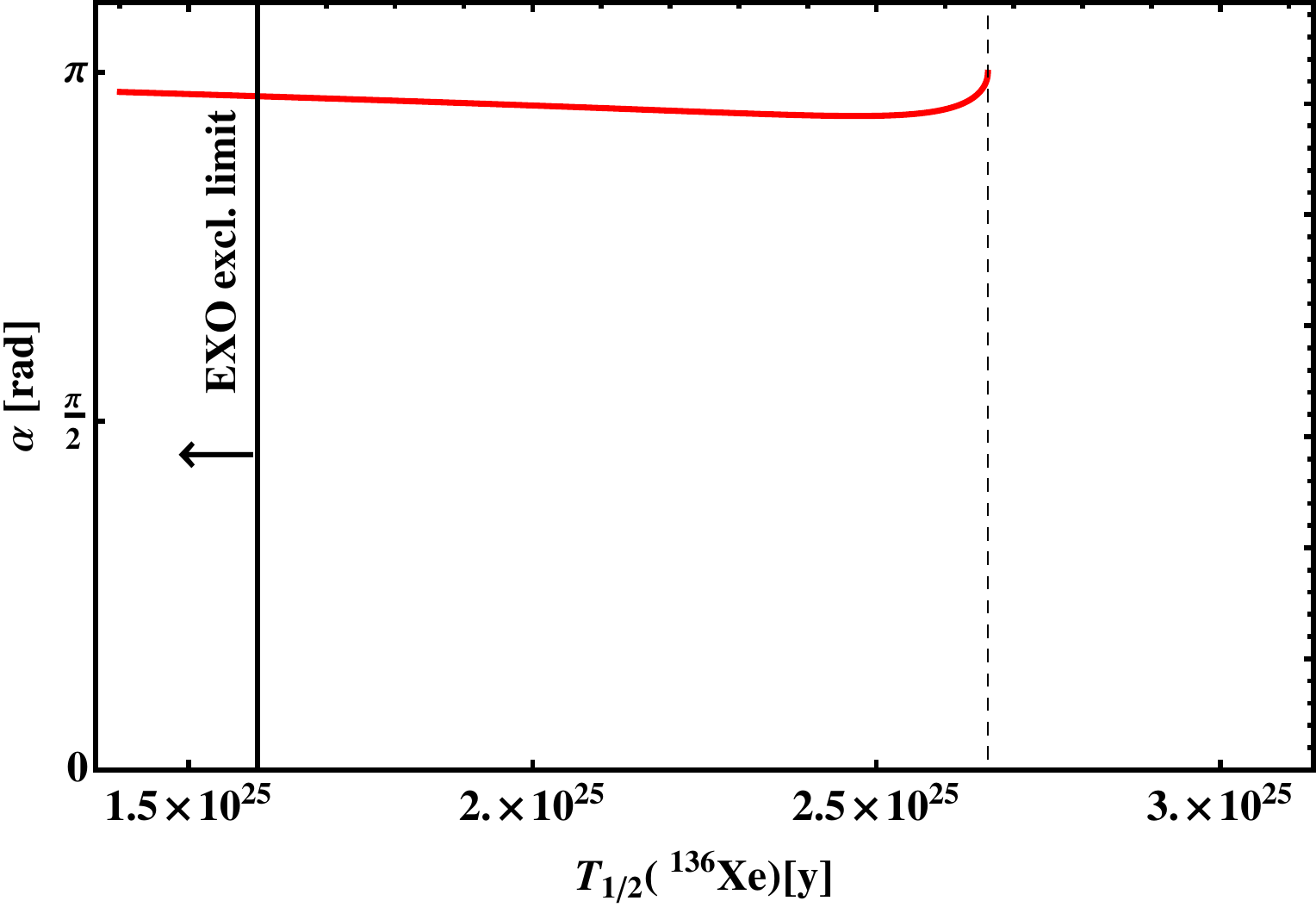}}
     \end{center}
\vspace{-1.0cm}
    \caption{
\label{fig:figInt1} Left panel: the values of $|\eta_\nu|^2\times
10^{10}$ (thick solid line) and $|\eta_{\lambda'}|^2\times 10^{14}$
(dotted line), obtained as solutions of the system of equations
(\ref{hlint}) for fixed values of $T^{0\nu}_{1/2}(^{76}$Ge$) = 2.23 \times 10^{25}$
y and $T^{0\nu}_{1/2}(^{130}$Te$) = 10^{25}$ y,  and letting
$T^{0\nu}_{1/2}(^{136}$Xe$)$ free. The physical allowed regions
correspond to the areas within the two vertical lines. Right panel:
the values of the phase $\alpha$ in the allowed interval of values
of $T^{0\nu}_{1/2}(^{136}$Xe$)$, corresponding to physical solutions
for  $|\eta_\nu|^2$ and $|\eta_{\lambda'}|^2$. In this case  $\cos
\alpha < 0$ and the interference is destructive. See text for
details.}
\end{figure}

\begin{figure}[h!]
  \begin{center}
 \subfigure
 {\includegraphics[width=7cm]{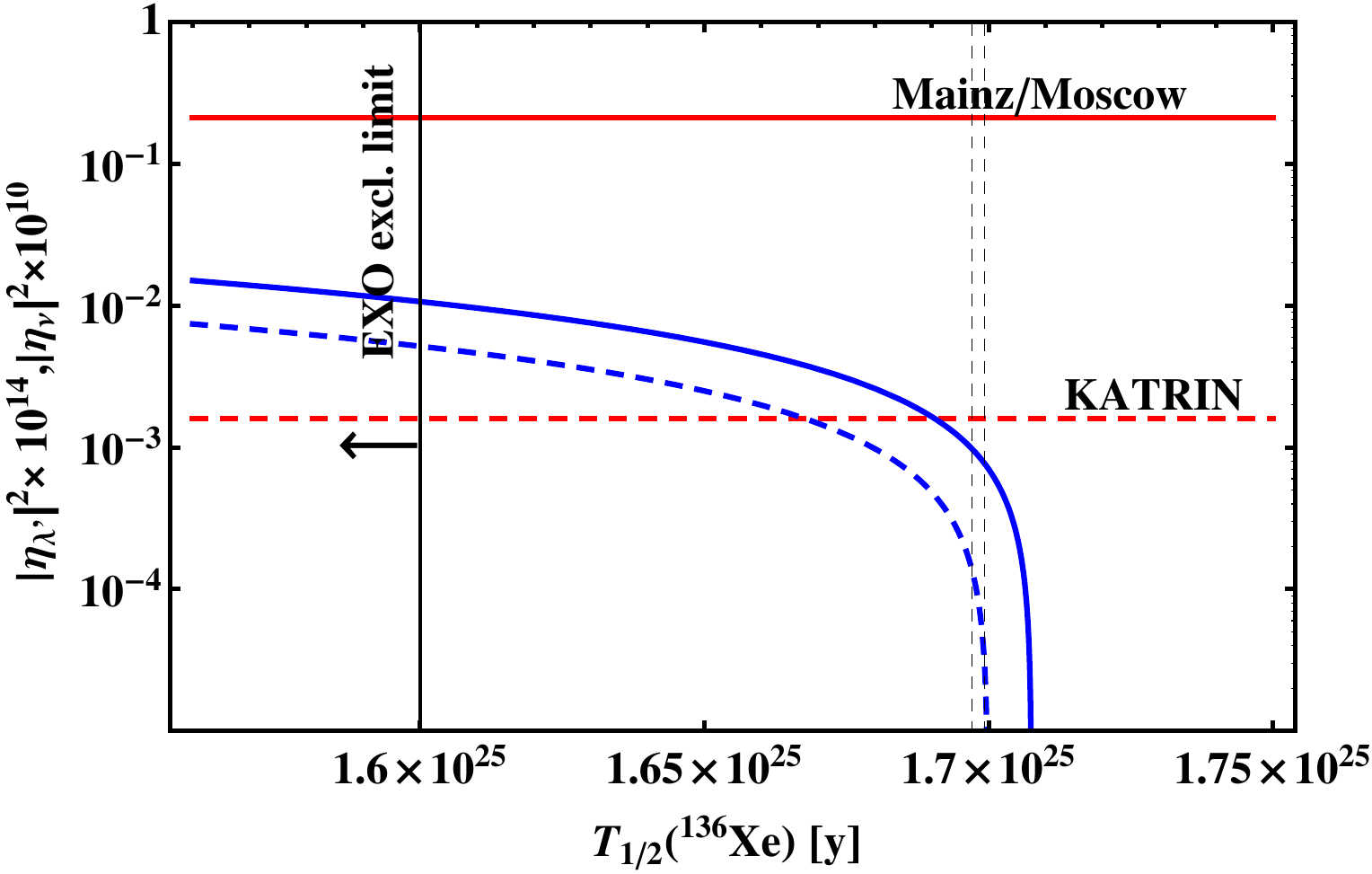}}
 \vspace{5mm}
 \subfigure
   {\includegraphics[width=6.3cm]{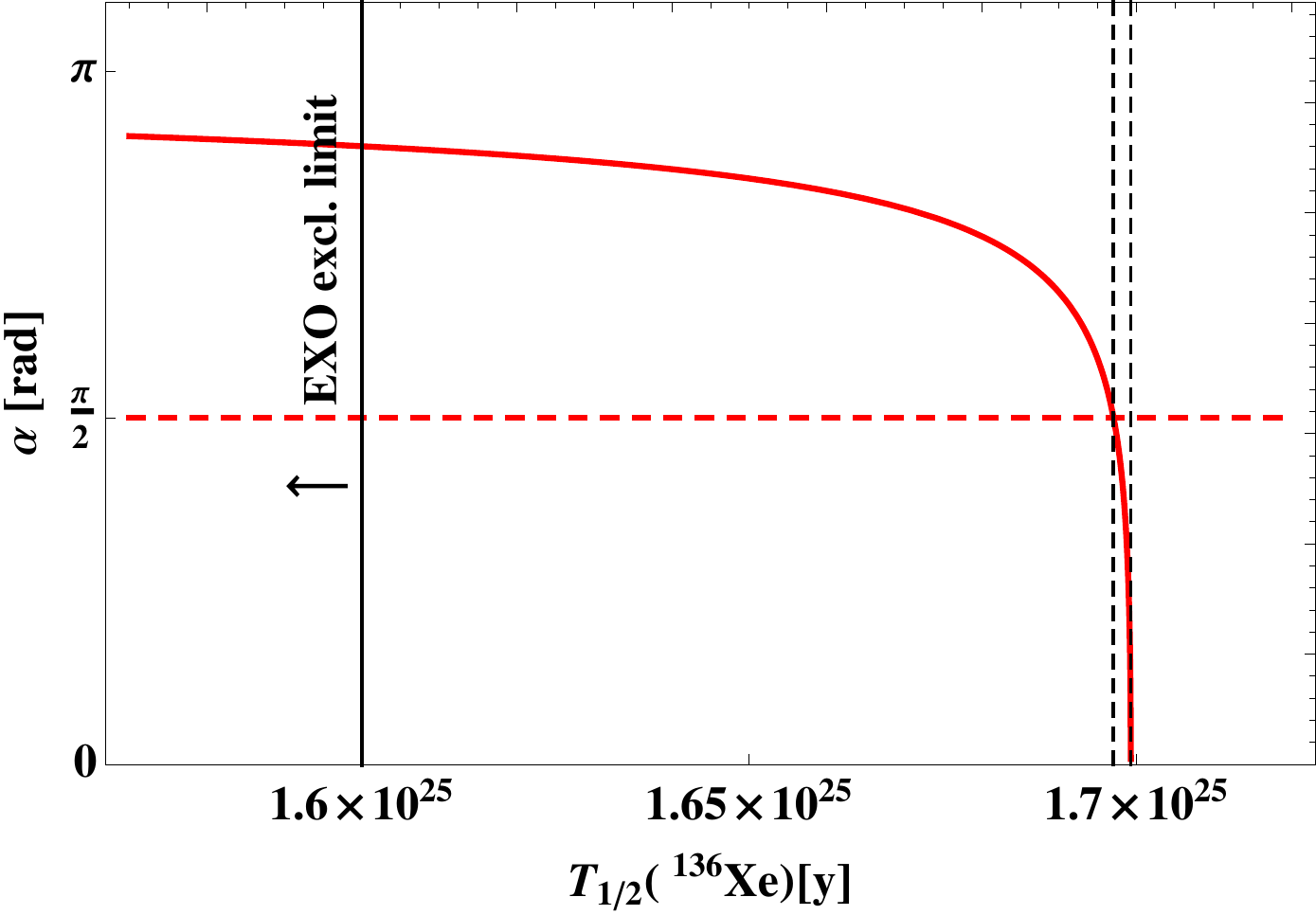}}
     \end{center}
\vspace{-1.0cm}
    \caption{
\label{fig:figInt2} Left panel: the same as in Fig.
\ref{fig:figInt1} but for $T^{0\nu}_{1/2}(^{76}$Ge$) = 3.5 \times 10^{25}$ y and
$T^{0\nu}_{1/2}(^{130}$Te$) = 8.0 \times 10^{25}$ y. The interval of values of
$T^{0\nu}_{1/2}(^{136}$Xe$)$ between i) the vertical solid and right
dashed lines ii) the two vertical dashed lines, and iii) the
vertical solid and left dashed lines, correspond respectively to i)
physical (non-negative) solutions for $|\eta_\nu|^2$ and
$|\eta_{\lambda'}|^2$,  ii) constructive interference ($z > 0$), and
iii) destructive interference ($z < 0$). Right panel: the
corresponding values of the phase $\alpha$ as a function of
$T^{0\nu}_{1/2}(^{136}$Xe$)$. Constructive interference is possible
only for values of $T^{0\nu}_{1/2}(^{136}$Xe$)$ between the two
vertical dashed lines. See text for details.}
\end{figure}

  The possibilities of destructive and constructive interference
are illustrated in Figs.  \ref{fig:figInt1} and \ref{fig:figInt2},
respectively. In these figures the physical allowed regions,
determined through the positivity conditions, correspond to the
areas within the two vertical lines (the solutions must be
compatible also with the existing lower limits given in eq
(\ref{limit})). For instance, using the Argonne ``large basis'' NMEs
corresponding to $g_A = 1.25$ and setting $T(^{76}Ge) = 2.23 \times
10^{25}$ y and $T(^{130}Te) = 10^{25}$ y, positive solutions are
allowed only in the  interval $1.60 \times 10^{25}\leq  T(^{136}Xe)
\leq 2.66 \times 10^{25}$ y (Fig. \ref{fig:figInt1}). As can be seen
in  Figs.  \ref{fig:figInt1} and \ref{fig:figInt2}, a constructive
interference is possible only if $T_2 \equiv T(^{136}$Xe$)$ lies in a
relatively narrow interval and $T_3\equiv T(^{130}$Te$)$ is determined
through the conditions in  eq. (\ref{constrint}).
%

\begin{figure}
  \begin{center}
 \subfigure
 {\includegraphics[width=7cm]{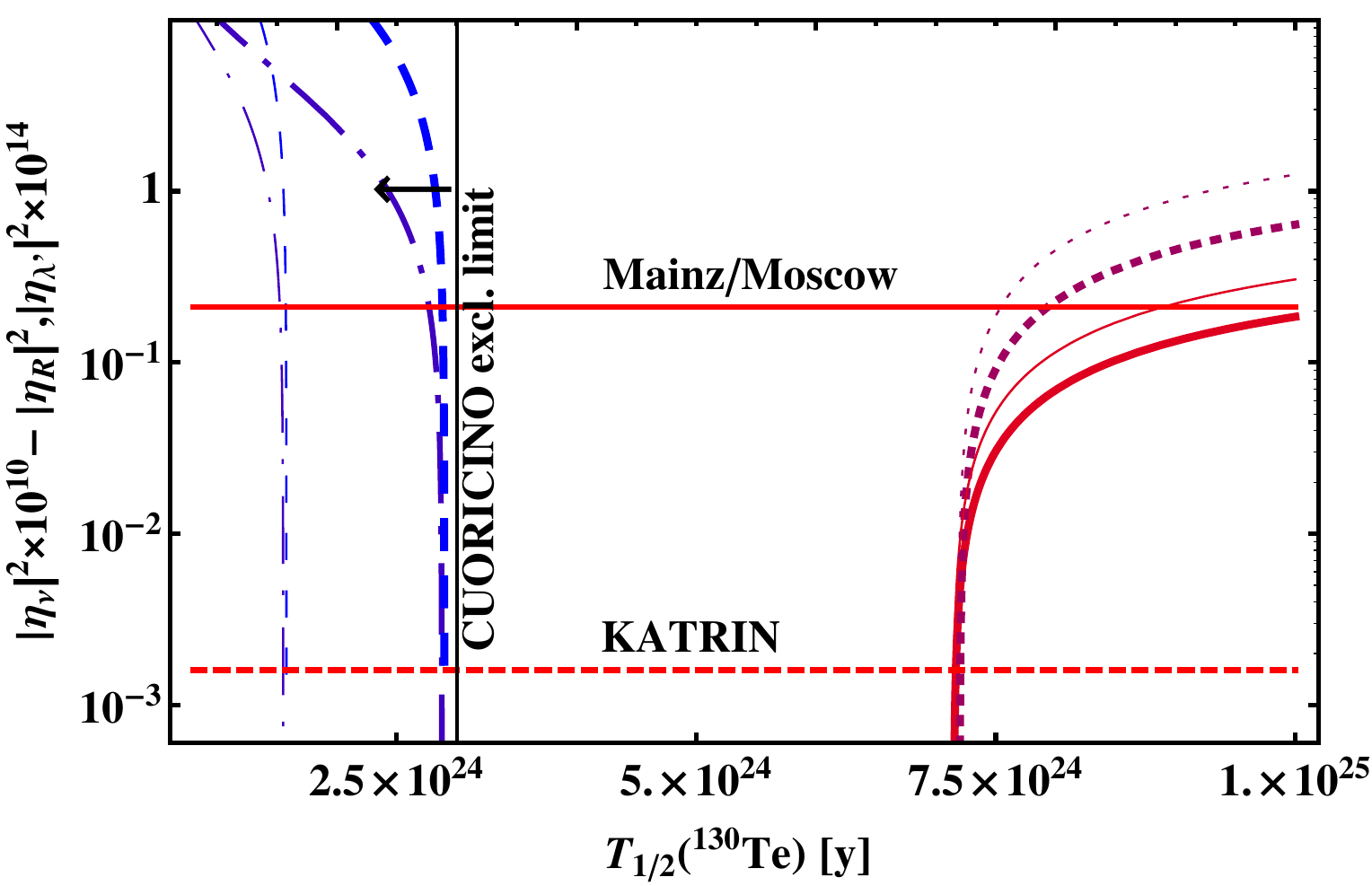}}
 \vspace{5mm}
 \subfigure
   {\includegraphics[width=7cm]{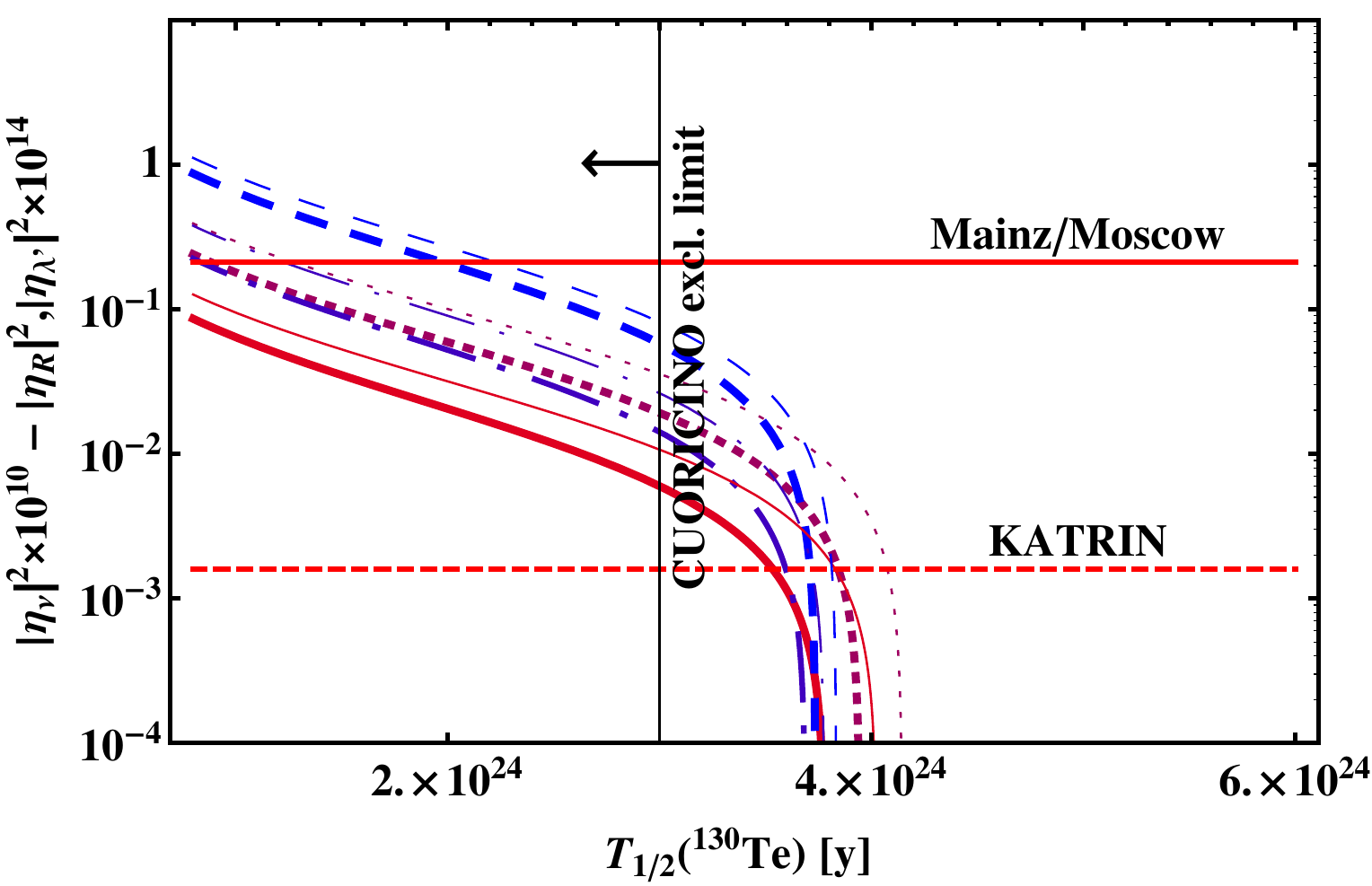}}
     \end{center}
\vspace{-1.0cm}
    \caption{\label{fig:figd1}
The parameters $|\eta_\nu|^2\times 10^{10}$ (solid line) and
$|\eta_{_L}|^2\times 10^{14}$ (dotted line) of the light and heavy LH
Majorana neutrino exchange mechanisms, and
$|\eta_{\lambda'}|^2\times 10^{14}$ (dashed-dotted line) and
$|\eta_{_L}|^2\times 10^{14}$ (dashed line) of the gluino and  heavy LH
Majorana neutrino exchange mechanisms, obtained from eq.
(\ref{intsol1}) using the Argonne NMEs (left panel) and CD-Bonn NMEs
(right panel), corresponding to $g_A=1.25$ (thick lines) and $g_A=1$
(thin lines), for  $T^{0\nu}_{1/2}(^{76}$Ge$)=2.23 \times 10^{25}$y,
$T^{0\nu}_{1/2}(^{136}$Xe$)=1.60 \times 10^{25}$y and letting
$T^{0\nu}_{1/2}(^{130}$Te$)$ free. See text for details.}
\end{figure}

 Next, we would like to illustrate the possibility to distinguish
between two pairs of interfering mechanisms i) A+B and ii) B+C,
which share one mechanism, namely B, from the data on the half-lives
of three isotopes. In this case we can set two systems of three
equations, each one in three unknowns. We will denote the
corresponding LNV parameters as i) $|\eta_A|^2$, $|\eta_B|^2$ and
ii) $|\eta_B|^2$ and $|\eta_C|^2$, while the interference parameters
will be denoted as i) $z$ and ii) $z'$. Fixing two of the three
half-lives, say $T_i$ and $T_j$, the possibility to discriminate
between the mechanisms $A$ and $C$ relies on the dependence of
$|\eta_A|^2$ and $|\eta_C|^2$ on the third half-life, $T_k$. Given
$T_i$ and $T_j$,  it will be possible to discriminate between the
mechanisms $A$ and $C$ if the two intervals of values of $T_k$ where
$|\eta_A|^2 > 0$ and $|\eta_C|^2 > 0$, do not overlap. If, instead,
the two intervals partially overlap, complete discrimination would
be impossible, but there would be a large interval of values of
$T_k$ (or equivalently, positive solutions values of the LNV
parameters) that can be excluded using present or future
experimental data. In order to have non-overlapping positive
solution intervals of $T_K$, corresponding to $|\eta_A|^2 > 0$ and
$|\eta_C|^2 > 0$, the following inequality must hold:
\be \frac{({\cal M}^{0\nu}_{k,A} {\cal M}^{0\nu}_{i,B}-{\cal M}^{0\nu}_{i,A}
{\cal M}^{0\nu}_{k,B}) ( {\cal M}^{0\nu}_{k,A}
{\cal M}^{0\nu}_{j,B}-{\cal M}^{0\nu}_{j,A} {\cal M}^{0\nu}_{k,B})}
{({\cal M}^{0\nu}_{k,B} {\cal M}^{0\nu}_{i,C}-{\cal M}^{0\nu}_{i,B}
{\cal M}^{0\nu}_{k,C}) ( {\cal M}^{0\nu}_{k,B}
{\cal M}^{0\nu}_{j,C}-{\cal M}^{0\nu}_{j,B} {\cal M}^{0\nu}_{k,C}) }<0.
\label{conddiscr} \ee
%
The above condition can be satisfied only for certain sets of
isotopes. Obviously, whether it is fulfilled or not depends on the
values of the relevant NMEs. We will illustrate this on the example
of an oversimplified analysis involving the light Majorana neutrino
exchange, the heavy LH Majorana neutrino exchange and the gluino
exchange as mechanisms $A$, $B$ and $C$, respectively, and the
half-lives of $^{76}$Ge, $^{130}$Te and  $^{136}$Xe:  $T_1 \equiv
T(^{76}Ge)$, $T_2\equiv T(^{130}Te)$ and  $T_3\equiv T(^{136}Xe)$.
Fixing  $T_1=  2.23 \times 10^{25}$ y and $T_3= 1.6 \times 10^{25}$
y (the EXO 90\% C.L. lower limit), we obtain the results shown  in
Fig. \ref{fig:figd1}. As it follows from Fig. \ref{fig:figd1}, in
the case of the Argonne NMEs (left panel), it is possible to
discriminate between the standard light neutrino exchange and the
gluino exchange mechanisms: the intervals of values of $T_2$, where
the positive solutions for the LNV parameters of the two pairs of
interfering mechanisms considered occur, do not overlap. Further,
the physical solutions for the two LNV parameters related to the
gluino mechanism are excluded by the CUORICINO limit on
$T(^{130}Te)$ \cite{Arnaboldi08}. This result does not change with the
increasing of $T_3$. Thus, we are lead to conclude that for $T_3$ $>$
1.6 $\times$ 10$^{25}$ y and $T_1$ given by the value claimed in
\cite{KlapdorK:2006ff}, of the two considered pairs of
possible interfering $\betabeta$-decay mechanisms, only the light
and heavy LH Majorana neutrino exchanges can be generating the
decay. The solution for $|\eta_\nu|^2$ must be compatible with the
upper limit $\meff < 2.3$ eV \cite{Lobashev,Eitel:2005hg}, indicated
with a solid horizontal line in Fig. \ref{fig:figd1}. In the right
panel of Fig. \ref{fig:figd1} we plot also the solutions obtained
with the  CD-Bonn NMEs. In this case is not possible to discriminate
between the two considered pair of mechanisms since the condition in
eq. (\ref{conddiscr}) is not satisfied.

\begin{figure}
  \begin{center}
 \subfigure
 {\includegraphics[width=7cm]{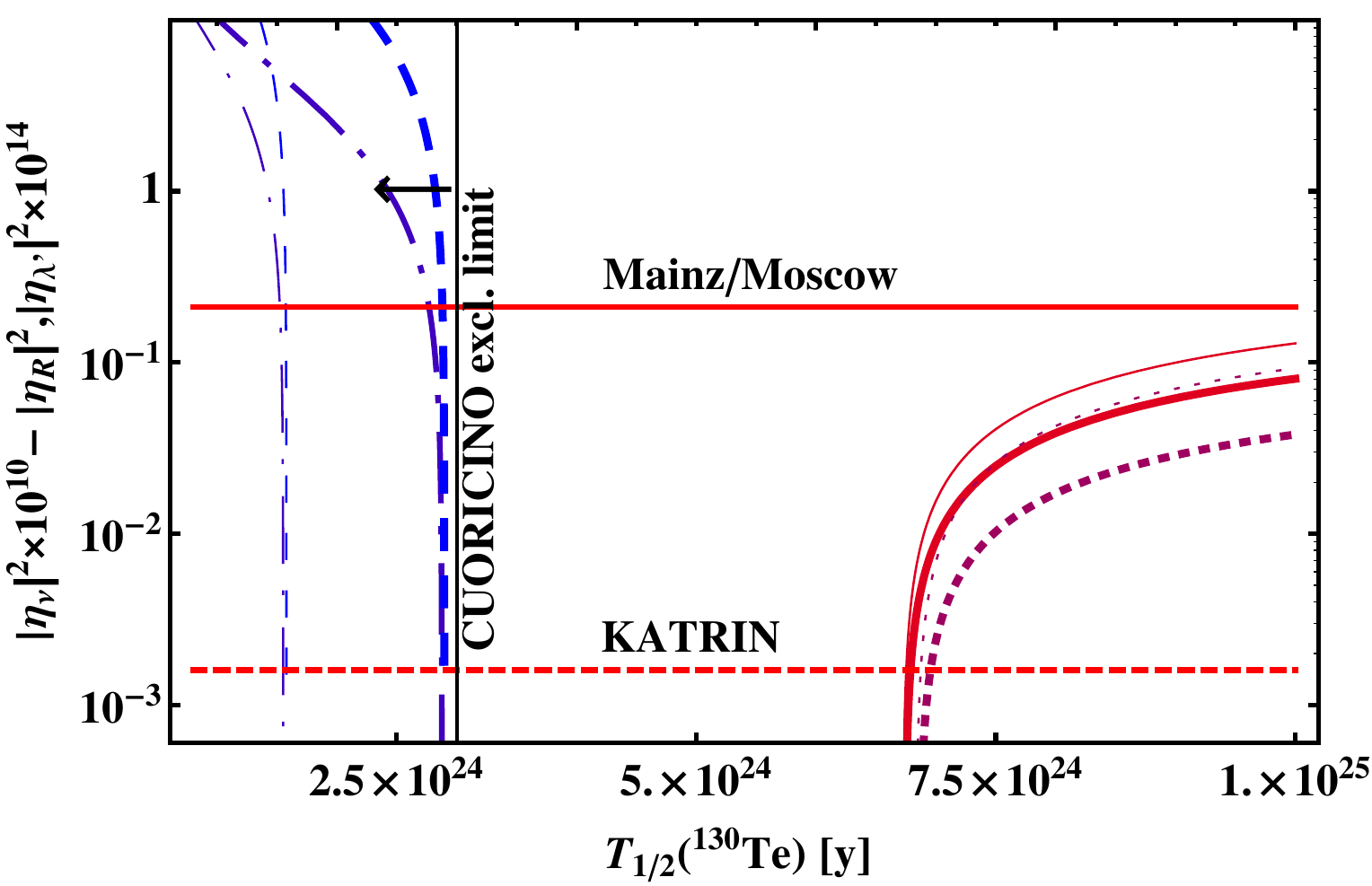}}
 \vspace{5mm}
 \subfigure
   {\includegraphics[width=7cm]{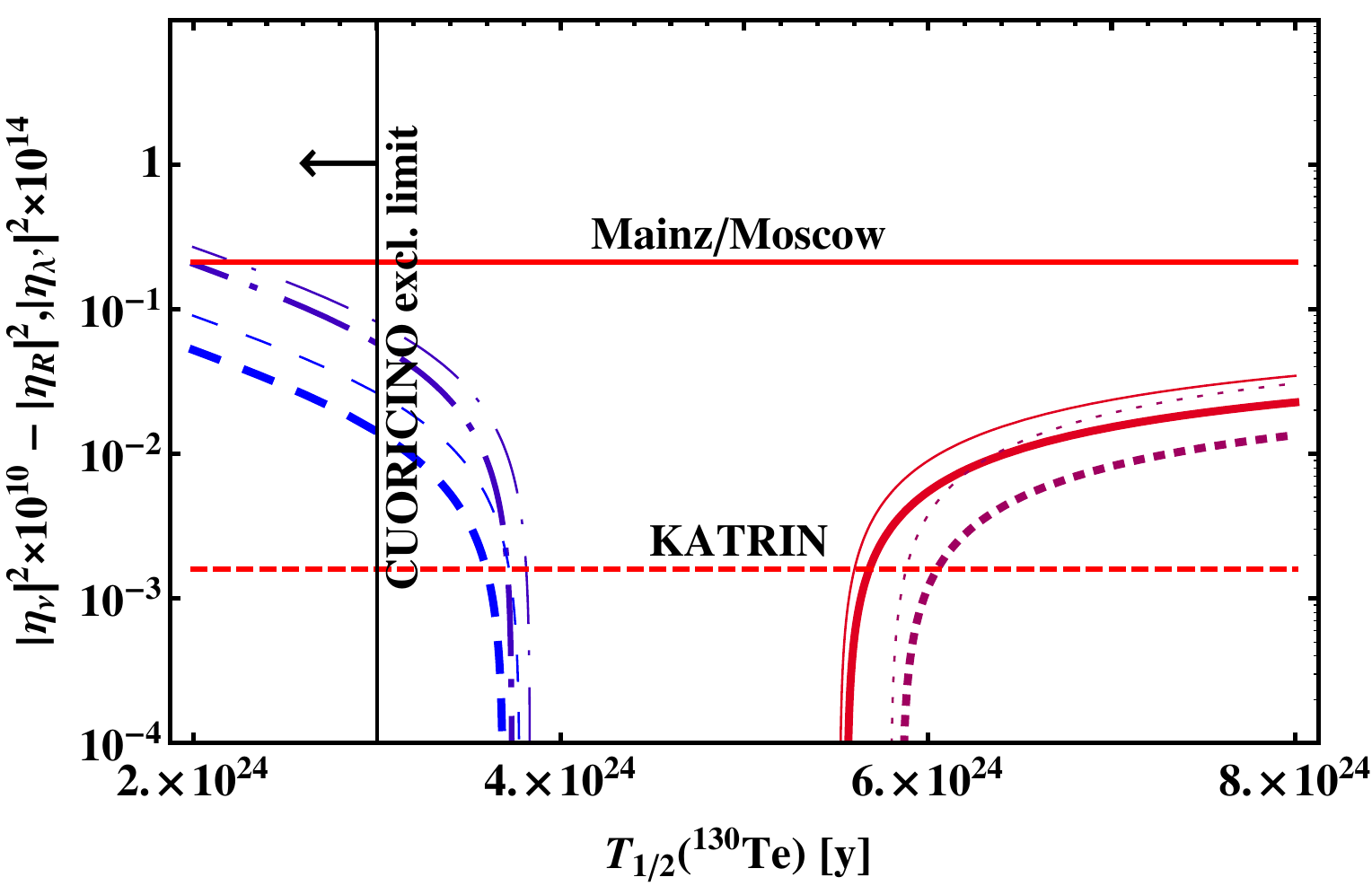}}
     \end{center}
\vspace{-1.0cm}
    \caption{\label{fig:figd2}
The same as in Fig. \ref{fig:figd1}, but for i) $|\eta_\nu|^2\times
10^{10}$ (thick solid line) and $|\eta_{\lambda'}|^2\times 10^{14}$
(thick dotted line) of the
 light neutrino and gluino exchange mechanisms, and
ii) $|\eta_{_L}|^2\times 10^{14}$ (thick dashed-dotted line) and
$|\eta_{\lambda'}|^2\times 10^{14}$ (thick dashed line) of the heavy
LH Majorana neutrino and gluino exchange mechanisms, and using
$T^{0\nu}_{1/2}(^{76}$Ge$)=2.23 \times 10^{25}$ y and
$T^{0\nu}_{1/2}(^{136}$Xe$)=1.60 \times 10^{25}$ y. See text for
details. }
\end{figure}

  Another interesting example is the case in which $A$ is the light
Majorana neutrino exchange, $B$ is the gluino exchange and $C$ the
heavy LH Majorana neutrino exchange, i.e., we try to discriminate
between i) the light neutrino plus gluino exchange mechanisms, and
ii) the heavy LH Majorana neutrino plus gluino exchange mechanisms.
We fix, like in the previous case, the values for $T_1=  2.23 \times
10^{25}$y and $T_3= 1.6\times 10^{25}$y. The results of this
analysis are plotted in Fig. \ref{fig:figd2}. Since the condition in
eq. (\ref{conddiscr}) is now satisfied for NMEs obtained either with
the Argonne potential or with the CD-Bonn potential, in this case it
is possible, in principle, to discriminate between the the two pair
of mechanisms independently of the set of NMEs used (within the sets
considered by us). This result does not change with the increasing
of $T_3$. Hence, as far as $T_1$ is fixed to the value claimed in
\cite{KlapdorK:2006ff} and the limits in eq.
(\ref{limit}) are satisfied, the two intervals of values of $T_2$,
in which the ``positivity conditions'' for i) $|\eta_\nu|^2$,
$|\eta_{\lambda'}|^2$ and $z$, and for ii) $|\eta_{\lambda'}|^2$,
$|\eta_N|^2$ and $z'$, are satisfied, are not overlapping  (Fig.
\ref{fig:figd2}).

%
\section{Final Remarks}
%

We have investigated the possibility to discriminate between
different pairs of CP non-conserving mechanisms inducing the
neutrinoless double beta $\betabeta$-decay by using a multi-isotope approach. 
In particular, data on
$\betabeta$-decay half-lives of nuclei with largely different
NMEs can be used to obtain information on couple of (non)-interfering mechanisms in \betabeta-decay. 
The mechanisms studied are: light
Majorana neutrino exchange, heavy left-handed (LH) and heavy
right-handed (RH) Majorana neutrino exchanges, lepton charge
non-conserving couplings in SUSY theories with $R$-parity breaking
giving rise to the ``dominant gluino exchange''.
 Each of the mechanisms is
characterized by a specific lepton number violating (LNV) parameter
$\eta_{\kappa}$, where the index $\kappa$ labels the mechanism. For
the five mechanisms listed above we use the notations $\kappa =
\nu,L,R,\lambda'$, respectively. The parameter
$\eta_{\kappa}$ will be complex, in general,  if the mechanism
$\kappa$ does not conserve the CP symmetry. The nuclei considered
are $^{76}$Ge, $^{82}$Se, $^{100}$Mo, $^{130}$Te and $^{136}$Xe.
\\
Four sets of NMEs of the
$\betabeta$-decays of these five nuclei, derived within the
Self-consistent Renormalized Quasiparticle Random Phase
Approximation (SRQRPA), were employed in our analysis. They
correspond to two types of nucleon-nucleon potentials - Argonne
(``Argonne NMEs'') and CD-Bonn (``CD-Bonn NMEs''), and two values of
the axial coupling constant $g_A = 1.25;1.00$. Given the NMEs and
the phase space factors of the decays, the half-life of a given
nucleus depends on the parameters  $|\eta_{\kappa}|^2$ of the
mechanisms triggering the decay.\\
We have considered in detail
the cases of two non-interfering and two interfering mechanisms
inducing the $\betabeta$-decay. If two non-interfering mechanisms
$A$ and $B$ cause the decay, the parameters $|\eta_{A}|^2$ and
$|\eta_{B}|^2$ can be determined from data on the half-lives of two
isotopes, $T_1$ and $T_2$ as solutions of a system of two linear
equations. If the half-life of one isotope is known, say $T_1$, the
positivity condition which the solutions $|\eta_{A}|^2$ and
$|\eta_{B}|^2$ must satisfy, $|\eta_{A}|^2 \ge 0$ and $|\eta_{B}|^2
\ge 0$, constrain the half-life of the second isotope $T_2$ (and the
half-life of any other isotope for that matter) to lie in a specific
interval. If $A$ and $B$ are interfering
mechanisms, $|\eta_{A}|^2$ and $|\eta_{B}|^2$ and the interference
term parameter, $z _{AB} \equiv 2\cos\alpha_{AB}|\eta_{A}\eta_{B}|$
which involves the cosine of an unknown relative phase $\alpha_{AB}$
of $\eta_{A}$ and $\eta_{B}$, can be uniquely determined, in
principle, from data on the half-lives of three nuclei, $T_{1,2,3}$.
In this case, given the half-life of one isotope, say $T_1$, the
``positivity conditions'' $|\eta_{A}|^2 \ge 0$, $|\eta_{B}|^2 \ge 0$
and $-1\leq \cos\alpha_{AB} \leq 1$ constrain the half-life of a
second isotope, say $T_2$, to lie in a specific interval, and the
half-life of a third one, $T_3$, to lie in an interval which is
determined by the value of $T_1$ and the interval of allowed values
of $T_2$.

  For all possible pairs of non-interfering mechanisms
we have considered these ``positivity condition''
intervals of values of $T_2$ were shown  to be
essentially degenerate if $T_1$ and $T_2$ correspond to the
half-lives of any pair of the four nuclei $^{76}$Ge, $^{82}$Se,
$^{100}$Mo and $^{130}$Te. This is a consequence of the fact that
for each of the five single mechanisms discussed, the NMEs for
$^{76}$Ge, $^{82}$Se, $^{100}$Mo and $^{130}$Te differ relatively
no more than 10\% \cite{FSV10MM,Faessler:2011qw}. One has
similar degeneracy of ``positivity condition'' intervals $T_2$ and
$T_3$ in the cases of two constructively interfering mechanisms
(within the set considered). These degeneracies might irreparably
plague the interpretation of the $\betabeta$-decay data if the
process will be observed.

  The NMEs for $^{136}Xe$,  resulting from calculations which
use the SRQRPA method, differ
significantly from those of $^{76}$Ge, $^{82}Se$, $^{100}$Mo and
$^{130}$Te, being by a factor $\sim (1.3 - 2.5)$ smaller. As we have
shown in the present section, this allows to lift to a certain
degree the indicated degeneracies and to draw conclusions about the
pair of non-interfering (interfering) mechanisms possibly inducing
the $\betabeta$-decay from data on the half-lives of $^{136}Xe$ and
of at least one (two) more isotope(s) which can be, e.g., any of the
four, $^{76}$Ge, $^{82}$Se, $^{100}$Mo and $^{130}$Te considered.

 We have analyzed also the possibility to discriminate
between two pairs of non-interfering (or interfering)
$\betabeta$-decay mechanisms when the pairs have one mechanism in
common, i.e., between the mechanisms i) $A$ + $B$ and ii) $C$ + $B$,
using the half-lives of the same two isotopes. We have derived the
general conditions under which it would be possible, in principle,
to identify which pair of mechanisms is inducing the decay (if any).
We have shown that the conditions of interest are fulfilled, e.g.,
for the following two pairs of non-interfering mechanisms i) light
neutrino exchange (A) and heavy RH Majorana neutrino exchange (B)
and ii) gluino exchange (C) and heavy RH Majorana neutrino exchange
(B), and for the following two pairs of interfering mechanisms i)
light neutrino exchange (A) and heavy LH Majorana neutrino exchange
(B) and ii) gluino exchange (C) and heavy LH Majorana neutrino
exchange (B), if one uses the Argonne NMEs in the analysis. They are
fulfilled for both the Argonne NMEs and CD-Bonn NMEs, e.g., for the
following two pairs of interfering mechanisms i) light neutrino
exchange (A) and gluino exchange (B), and ii) heavy LH Majorana
neutrino exchange (C) and  gluino exchange (B).

The  results obtained  here show that using the $\betabeta$-decay half-lives of
nuclei with largely different NMEs would help
resolving the problem of identifying the mechanisms triggering the
decay.

Concluding, 
 an eventual observation of a Majorana field would be a fundamental headway. This would mean that Nature admits the existence of particles which are identical to their anti-particles and, more importantly, it could point to the existence of New Physics, or in other words to new lepton number violating couplings in the Lagrangian of particle interactions. The data on \betabeta-decay, which will be available from the currently running experiments GERDA, EXO-200, KamLAND-Zen and from the CUORE experiment will be of crucial importance to identify the mechanism(s) triggering the decay if the latter will be observed. This will help to identify the New Physics beyond that predicted by the Standard Model associated with lepton charge non-conservation and the \betabeta-decay.

\bibliographystyle{unsrt}
 \bibliography{biblio}

\end{document}